\documentclass[prl,twocolumn,eqsecnum,superscriptaddress]{revtex4-1}
\usepackage{amsmath,amssymb,amsfonts,graphicx,color,verbatim,hyperref,fancyref,mathrsfs}

\newcommand{\beq}{\begin{equation}}
\newcommand{\eeq}{\end{equation}}
\def\bea{\begin{eqnarray}}
\def\eea{\end{eqnarray}}

\newcommand{\ket}[1]{\left| #1 \right\rangle }
\newcommand{\bra}[1]{\left\langle #1 \right|}
\newcommand{\avg}[1]{\langle #1 \rangle}
\newcommand{\abs}[1]{\vert #1 \vert}
\newcommand{\adj}[1]{#1^{\dagger}}

\def\({\left(}
\def\){\right)}

\def\CC{{\cal C}}

\def\CH{{\cal H}}

\def\CL{{\cal L}}

\def\CS{{\cal S}}

\begin{document}

\title{Measuring the scrambling of quantum information}


\author{Brian Swingle}
\email{bswingle@stanford.edu}
\affiliation{Department of Physics, Stanford University, Stanford, California 94305, USA}
\affiliation{Stanford Institute for Theoretical Physics, Stanford, California 94305, USA}

\author{Gregory Bentsen}
\affiliation{Department of Physics, Stanford University, Stanford, California 94305, USA}

\author{Monika Schleier-Smith}
\affiliation{Department of Physics, Stanford University, Stanford, California 94305, USA}

\author{Patrick Hayden}
\affiliation{Department of Physics, Stanford University, Stanford, California 94305, USA}
\affiliation{Stanford Institute for Theoretical Physics, Stanford, California 94305, USA}

\begin{abstract}
We provide a protocol to measure out-of-time-order correlation functions. These correlation functions are of theoretical interest for diagnosing the scrambling of quantum information in black holes and strongly interacting quantum systems generally.  Measuring them requires an echo-type sequence in which the sign of a many-body Hamiltonian is reversed.  We detail an implementation employing cold atoms and cavity quantum electrodynamics to realize the chaotic kicked top model, and we analyze effects of dissipation to verify its feasibility with current technology. Finally, we sketch prospects for measuring out-of-time-order correlation functions in other experimental platforms.
\end{abstract}

\maketitle

Advances in the coherent manipulation of quantum many-body systems are enabling measurements of the dynamics of quantum information \cite{Cheneau12,Richerme14,Jurcevic14,Eisert15}. Notably, recent experiments \cite{Cheneau12} have corroborated the Lieb-Robinson bound, a fundamental speed limit on the propagation of signals even in non-relativistic spin systems \cite{lieb72finite}.  At the same time, new theoretical bounds have been derived from the study of black holes \cite{BHandBF-SS2013}.  Consistent with their wide variety of extreme physical properties, black holes saturate several absolute limits on quantum information processing. They are the densest memories in nature \cite{Bekenstein:1973ur}.  They also process their information extremely rapidly \cite{haydenpreskill,fastscramble} and reach a conjectured bound on the rate of growth of chaos \cite{chaosbound}.

That black holes process quantum information at all is demonstrated by the holographic principle~\cite{hooft1993dimensional,susskind1995world}: a black hole in anti-de Sitter space is equivalent to a thermal state of a lower dimensional quantum field theory \textit{without gravity} \cite{adscft}. This means that certain quantum mechanical systems \cite{bfss,magoo} that might in principle be realizable in experiments \cite{cold_atoms_gauge} are dynamically equivalent to black holes in quantum gravity.  A major open question is the extent to which familiar quantum many-body systems also behave like black holes. Besides potentially enabling experimental tests of the holographic principle, addressing this question will shed light on fundamental limits on quantum information processing.

For a quantum field theory to be the holographic dual of a black hole, its dynamics must be highly chaotic \cite{chaosbound}. To quantify the rate of growth of chaos, recent work has explored an inherently quantum mechanical version of the butterfly effect, namely, the growth of the commutator between two operators as a function of their separation in time \cite{BHandBF-SS2013}. This growth is indicative of a process known as scrambling \cite{BHandBF-SS2013,haydenpreskill,fastscramble,Hosur2016}, wherein a localized perturbation spreads across a quantum many-body system's degrees of freedom, thereby becoming inaccessible to local measurements. The timescale for scrambling is theoretically distinct from the relaxation time and has yet to be probed in any experiment.

In this Letter, we propose a broadly applicable protocol for measuring scrambling.  We describe a concrete implementation with cold atoms coupled to an optical cavity, a versatile platform for engineering spin models with non-local interactions \cite{sorensen02,leroux2010implementation,multimode,subircavity1,hosten2016quantum} that have the potential to exhibit fast scrambling dynamics at or near the chaos bound \cite{haydenpreskill,Lashkari13}.  We show that a realistic measurement---including coupling to the environment---can distinguish between time-scales for relaxation and scrambling in a globally interacting chaotic ``kicked top'' model, a special case of the non-local models accessible in the cavity-QED setting.  We also discuss prospects for measuring scrambling in local Hamiltonians using trapped ions, Rydberg atoms, or ultracold atoms in optical lattices.

We emphasize the generality of our approach because scrambling, while hitherto difficult to study outside the black-hole context, is of broad importance in quantum many-body dynamics.  Probing scrambling in diverse physical systems could elucidate links between chaos and fast computation \cite{Brown2016}, reveal novel ways of robustly hiding quantum information in non-local degrees of freedom \cite{Hosur2016}, uncover new bounds on transport coefficients \cite{2014PhRvE..89a2142P,2016arXiv160309298R,2016arXiv160308510B}, and offer insight into closed-system thermalization.  While identifying mechanisms that promote scrambling could aid in designing many-body systems dual to black holes, measuring scrambling may also illuminate under what conditions holographic duality is useful for understanding more conventional many-body systems.

We access the scrambling time via the decay of an out-of-time-order (OTO) correlation function \cite{BHandBF-SS2013,kitaevbhchaos}
\begin{equation}
F(t) = \langle  W_t^\dagger V^\dagger W_t V\rangle,
\end{equation}
where $V$ and $W$ are commuting unitary operators and $W_t = U(-t) W U(t)$ is the Heisenberg operator obtained by time evolution  $U(t)=e^{-i H t}$ (setting $\hbar=1$) under a Hamiltonian $H$.  Physically, $F$ describes a gedankenexperiment in which we are able to reverse the flow of time. We compare two quantum states obtained by either (1) applying $V$, waiting for a time $t$, and then applying $W$; or (2) applying $W$ at time $t$, going back in time to apply $V$ at $t=0$, and then letting time resume its forward progression to $t$ \cite{shenker2014multiple,roberts2015localized}. The correlator $F$ measures the overlap between the two final states.  In a many-body system with a nontrivial interaction Hamiltonian $H$, $F(t)$ diagnoses the spread of quantum information by measuring how quickly the interactions cause initially commuting operators $V$ and $W$ to fail to commute: $\langle \abs{[W_t,V]}^2\rangle = 2(1 - \text{Re}[F])$.

Because $F$ is always one in the absence of noncommutativity, it may be regarded as an intrinsically quantum mechanical variant of the Loschmidt echo \cite{EchoReview}, a paradigmatic probe of chaos.  The Loschmidt echo, which depends on time-ordered correlation functions, has been measured in several pioneering experiments \cite{EchoReview,HahnEcho,MagicEcho,PolarizationEcho,AtomEcho,gorin2006dynamics}.  Its decay can be related to the mean Lyapunov exponent of a corresponding chaotic classical system and to decoherence~\cite{DecoherenceChaosSecondLaw,DecoherenceJalabert,DecoherenceEchoZurek}.  By comparison, the decay rates of OTO correlators depend not only on Lyapunov exponents \cite{quasiclassicalsc} but also on the number of degrees of freedom: the higher the entropy, the slower the decay.

The growth of the scrambling time with entropy $\mathbb{S}$ can be understood from a model of $\mathbb{S}$ qubits (spins) evolving unitarily in discrete time \cite{Lashkari13}.  Intuitively, if we allow arbitrary pairwise interactions, the fastest way to delocalize information is to apply random two-body unitaries between $\mathbb{S}/2$ random pairs of spins at each time step of length $\tau$.  A single time-step suffices to make simple time-ordered autocorrelation functions decay, i.e., $\tau$ is the relaxation time.  Scrambling, however, requires information in one spin to spread to all the spins.  With information spreading exponentially fast to $2$, $4$, ..., $2^{t/\tau}$ spins, the scrambling time is then $t_* = \tau \log_2 \mathbb{S}$.

The relevance of the OTO correlator $F$ for accessing the scrambling time \cite{BHandBF-SS2013,kitaevbhchaos,chaosbound,Hosur2016} can be seen from the same random-circuit model.  For operators $W$ and $V$ that perturb individual spins, the typical time for $\langle \abs{[W_t,V]}^2\rangle$ to become order unity is $t_*$ because $W_t$ is supported on approximately $2^{t/\tau}$ spins.  Similarly, for black holes in Einstein gravity, scrambling occurs exponentially fast and is accompanied by an initial growth $1- F \sim e^{t/\tau}/\mathbb{S} + ...$, where the relaxation time $\tau=1/(2\pi T)$ is set by the temperature $T$ \cite{foot1}. The resulting decay time $t_* = \tau \ln \mathbb{S}$ is conjectured to be a fundamental bound for thermal states of time-independent Hamiltonians \cite{chaosbound}. Identifying bounds on scrambling in time-dependent models or in non-thermal states is an open problem, which experimental study of OTO correlators will help to address.

\begin{figure}
\includegraphics[width=.45\textwidth]{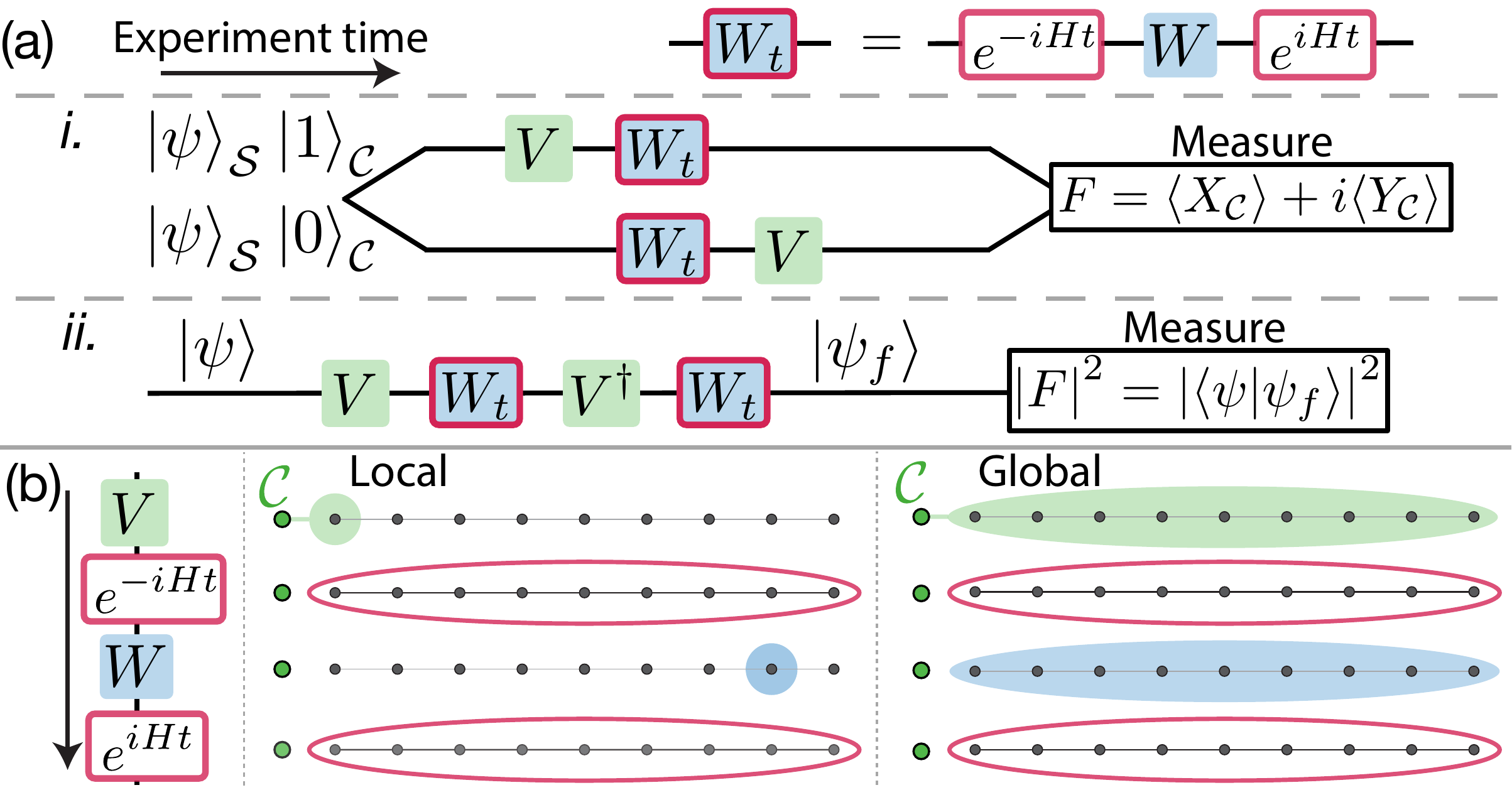}
\caption{(a) Protocols for measuring $F(t)$. (i) Given a control qubit $\mathcal{C}$, the \textit{interferometric protocol} can measure $F$ for the system $\mathcal{S}$ by applying different sequences of operators in the two interferometer arms. (ii) Without a control qubit, the \textit{distinguishability protocol} can access $|F|^2$. (b) $F(t)$ can be measured with either local or global operators $V, W$, as shown for a spin chain in the upper branch of the interferometer, with control qubit $\CC$ in green (left of chain).} 
\label{fig:circuit}
\end{figure}

A key capability required to measure the OTO correlator $F(t) = \langle  W_t^\dagger V^\dagger W_t V \rangle$ is that of reversing the sign of the Hamiltonian.  Obtaining full information about $F(t)$ additionally requires many-body interferometry, similar to schemes in Refs. \cite{Knap12,PhysRevLett.115.135302,PhysRevLett.113.020505,2014PhRvL.113b0505P,abanin2012measuring,rydberggate}.  A control qubit can be used to produce two branches in the many-body state \cite{abanin2012measuring,rydberggate,cavitygate}; measurements of the control qubit then reveal $F$ (Fig. \ref{fig:circuit}a). Even without the control qubit, an alternative protocol (Figure \ref{fig:circuit}b) suffices to measure the magnitude of $F$, which quantifies the indistinguishability of two states obtained by applying $V$ and $W_t$ in differing order.

As perfect time-reversal is impossible in practice, our protocol is the experimentally reasonable one: the Hamiltonian dynamics is reversed but dissipation is not.  It is thus important to establish that observables obtained from this partial time reversal access the same physics as the analogous observables in the unitary protocol.  We show that such a regime is achievable in the cavity model with current technology.

\textit{General protocol.} Consider a quantum system $\CS$ initialized in state $|\psi\rangle_{\CS}$.  Our goal is to measure the four point function $F(t) = \langle  W_t^\dagger V^\dagger W_t V \rangle$,
where $V$ and $W$ are simple unitary operators acting on $\CS$ which initially commute.  For $F$ to be non-trivial, the time evolution $U(t) = e^{-iHt}$ must be governed by a many-body Hamiltonian $H$ containing interactions between different degrees of freedom.  The Heisenberg operator $W_t= U(-t) W U(t)$ then grows in complexity as $t$ increases and eventually fails to commute with $V$.

The \textit{interferometric protocol} for measuring $F$ employs a control qubit $\CC$ initialized in state $\ket{+X}_\CC=\frac{|0\rangle_{\CC} + |1\rangle_{\CC}}{\sqrt{2}}$. First apply the gate sequence (illustrated in Figure \ref{fig:circuit}a)
\begin{eqnarray}
  &[1]:& \,I_{\CS} \otimes \ket{0}\bra{0}_{\CC} + V_{\CS} \otimes \ket{1}\bra{1}_{\CC} \nonumber \\
  &[2]:& \,U(t)_{\CS}\otimes I_{\CC} \nonumber \\
  &[3]:& \,W_{\CS} \otimes I_{\CC} \nonumber \\
  &[4]:& \,U(-t)_{\CS} \otimes I_{\CC} \nonumber \\
  &[5]:& \,V_{\CS} \otimes \ket{0}\bra{0}_{\CC} + I_{\CS} \otimes \ket{1}\bra{1}_{\CC} \nonumber
\end{eqnarray}

to prepare the state
\beq
\frac{( V W_t \ket{\psi}_{\CS} )\ket{0}_{\CC} + (W_t  V \ket{\psi}_{\CS} )\ket{1}_{\CC}}{\sqrt{2}}.
\eeq
Then measure the control qubit in the $X$ and $Y$ bases to find the real and imaginary parts of the OTO correlator
\begin{equation}
F = \avg{X_\CC} + i \avg{Y_\CC},
\end{equation}
where $X_\CC$ and $Y_\CC$ denote Pauli matrices acting on $\CC$.


Even without a control qubit, it is possible to measure the magnitude of $F$ using the \textit{distinguishability protocol}.  Initialize the system into state $\ket{\psi}$. Apply the gate sequence shown in Fig. \ref{fig:circuit}b to prepare the state $\ket{\psi_f} = W_t^\dagger V^\dagger W_t V \ket{\psi}$.  Finally, measure the projector $\Pi = \ket{\psi}\bra{\psi}$.  The result is
$\bra{\psi_f} \Pi \ket{\psi_f} = |F|^2$, which
quantifies the distinguishability of the two branches and is expected to contain roughly the same timescales as $F$.  As the projection $\Pi$ onto an arbitrary many-body state can be challenging to implement, the distinguishability protocol requires a careful choice of the initial state $\ket{\psi}$.

\textit{Cavity QED proposal.}  As a representative system amenable to probing the OTO correlator, we consider a collection of two-level atoms (spins) that interact via their mutual coupling to one or more modes of an optical cavity (Fig. \ref{fig:cavity_setup}). A drive laser incident from the side of the cavity generates interactions among all pairs of atoms it addresses. The sign of the interactions is set by the laser frequency, enabling access to the magnitudes of OTO correlators via the distinguishability protocol. To access the phase, a single individually addressable atom can serve as a control qubit for interferometry \cite{cavitygate}.

\begin{figure}
\includegraphics[width=.45\textwidth]{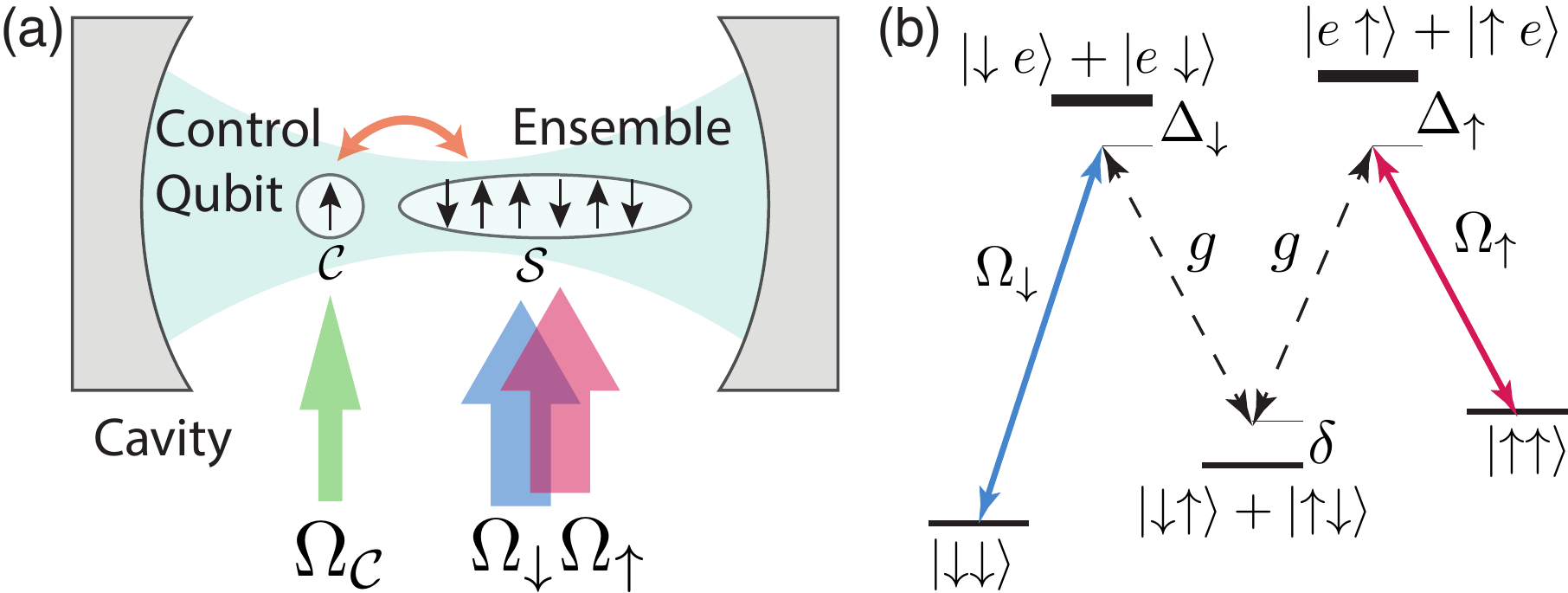}
\caption{Scheme for measuring out-of-time-order correlators. (a) Atomic ensemble $\CS$ and control qubit $\CC$ in an optical cavity are driven by control fields $\Omega_{\uparrow}, \Omega_{\downarrow}, \Omega_{\CC}$. (b) Control fields $\Omega_{\uparrow,\downarrow}$ and cavity coupling $g$ mediate pairwise interactions in the ensemble $\CS$ via 4-photon Raman transitions.}
\label{fig:cavity_setup}
\end{figure}

The cavity-mediated interactions within the ensemble generically take the form of a nonlocal spin model \cite{sorensen02,multimode,subircavity1}
\beq\label{eq:H}
H = \sum_{i j} J_{ij} s_i^x s_j^x + h.c.,
\eeq
where $\mathbf{s}_i$ is a pseudo-spin operator representing two internal atomic states (e.g., hyperfine states) $\ket{s_i^z=\pm 1/2}$. For $N$ atoms at positions $r_i$ with couplings $g_\alpha(r_i)$ to a set of degenerate cavity modes indexed by $\alpha$, the spin-spin couplings are given by
\beq\label{eq:Jij}
J_{ij} = \sum_\alpha \frac{\Omega_\uparrow^*(r_i)\Omega_\downarrow(r_j)}{\Delta_\uparrow\Delta_\downarrow} \frac{g_\alpha(r_i) g^*_\alpha(r_j) }{\delta},
\eeq
where $\Omega_{\uparrow,\downarrow}$ are the Rabi frequencies of the drive fields, detuned by $\Delta_{\uparrow,\downarrow}$ from atomic resonance, and $\delta$ is the detuning of the two-photon transition mediated by the drive fields and cavity couplings $g_{\alpha}$.

Key features of the light-mediated interactions are that their sign is controllable via the two-photon detuning $\delta$ \cite{Davis15}, they can be switched on and off, and the full graph of interactions can depend on the atomic positions and the spatial structure of the cavity modes and control fields. Also, it is possible to produce noncommuting $s^+ s^-$ type interactions, to add fields in any direction, and to include time dependence in the Hamiltonian. This versatility allows for studying a range of many-body phenomena, from quantum glasses \cite{sachdevye,multimode,subircavity1} to random circuit models that mimic the fast scrambling of black holes \cite{Lashkari13}.

\begin{figure}
\includegraphics[width=.45\textwidth]{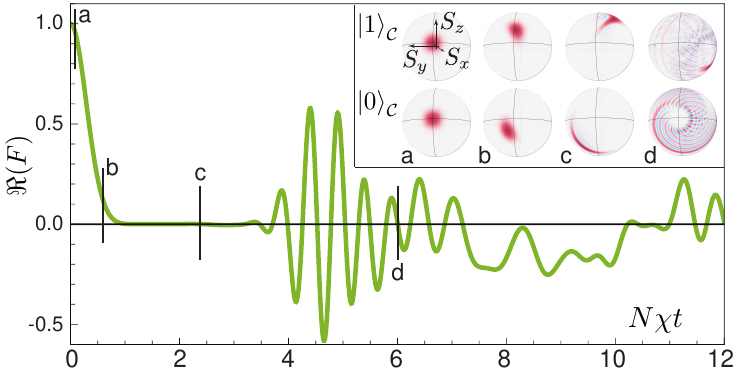}
\caption{Interferometric protocol for unitary $S_x^2$ dynamics at $N = 50$. For an initial coherent state $\left| \hat{y} \right\rangle$ and rotation angle $\phi = \pi / 4$, $\text{Re}[F]$ (green) exhibits decay at short times (a,b), a quiescent period (c), and subsequent oscillations (d). Inset: states of the two interferometer arms at various times, illustrated by Wigner quasiprobability distributions.}
\label{fig:szsq_blochspheres}
\end{figure}

For ease of visualization, we focus here on globally interacting spin models obtained by coupling all atoms uniformly to a single cavity mode.  Here, the Hamiltonian of Eq. \ref{eq:H} reduces to a ``one-axis twisting'' Hamiltonian $H_{\mathrm{twist}} = \chi S_x^2$, where $\mathbf{S} = \sum_i \mathbf{s}_i$ and the total spin is $S = N / 2$.   By considering correlators where the operations $V$ and $W$ are global spin rotations, we restrict the dynamics to a space of permutation-symmetric states that are conveniently described by quasiprobability distributions on a Bloch sphere (Fig. \ref{fig:szsq_blochspheres}).

To perform the controlled-$V$ step in the interferometer of Fig. \ref{fig:circuit}a, we convert the control qubit state $\ket{n}_{\CC}$ (with $n\in \{0,1\}$) into an $n$-photon state of the cavity, which produces a differential a.c. Stark shift $\propto n$ between each of the ensemble atoms' two levels. The result is a collective controlled phase gate
\beq
Z_\phi^{\CC} = I_{\CS}\otimes\ket{0}\bra{0}_{\CC} + e^{-i\phi S_z^{\CS}}\otimes \ket{1}\bra{1}_{\CC},
\eeq
where $S_z^{\CS} = \sum_i s_i^z$.  The rotation $W$, by contrast, is applied irrespective of the control qubit state.

Figure \ref{fig:szsq_blochspheres} shows calculated results of the interferometric protocol for the one-axis twisting model with an initial state $\ket{\hat{y}} = \ket{S_y = S}$ and rotations $V = W= e^{-i\phi S_z}$ with $\phi = \pi/4$.  Such a large controlled rotation is neither necessary nor sustainable at higher atom numbers, as discussed below, but it illustrates in exaggerated form the processes controlling $F$.  The initial decay of $F$ corresponds to the collective spin's trajectory on the Bloch sphere diverging between the two interferometer arms, as shown by Wigner quasiprobabilty distributions \cite{Dowling94} in Fig. \ref{fig:szsq_blochspheres}.  Later fluctuations in $F$ correspond to the spread of the quantum state over the entire Hilbert space.  


To further illuminate the physics of the OTO correlator, the $S_x^2$ Hamiltonian may be modified to produce a chaotic system.  Periodically applying a rapid $S_z$ rotation produces a ``kicked top'' model that has been studied both theoretically and experimentally \cite{haake,2009Natur.461..768C,2004PhRvE..70a6217W}. The stroboscopic dynamics are described by repeated application of the unitary operator $U = e^{- i k S_x^2/(2S)} e^{- i p S_z}$,
where $k$ measures the strength of interactions and $p$ measures the size of the rotational kick. Following Haake et al. \cite{haake}, we set $p=\pi/2$; then the corresponding classical model describes motion on the Bloch sphere which is regular for small $k$ and chaotic for large $k$. The semi-classical limit is the limit of large $S$, whereas previous experimental work has studied the case $S=3$ \cite{2009Natur.461..768C}.  The cavity-QED implementation proposed here, where the spin is scalable from the small-$S$ quantum regime to the semi-classical limit, provides an ideal testbed for probing the physics of the OTO correlator in a paradigmatic chaotic system.

\begin{figure}
\includegraphics[width=.45\textwidth]{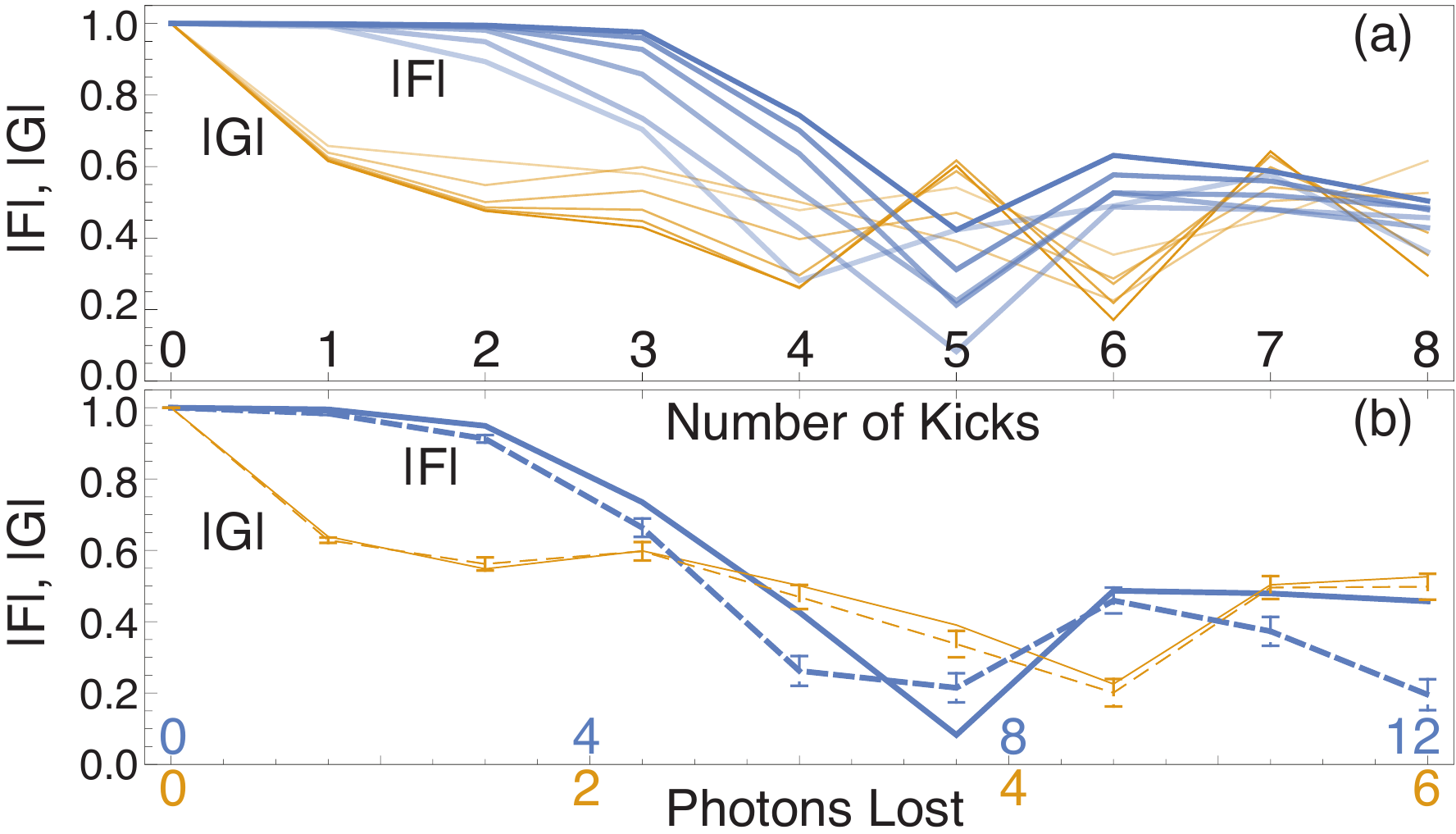}
\caption{Interferometric protocol for the kicked top. (a) Unitary time-ordered correlators $\abs{G(t)}$ (thin yellow) and out-of-time-order correlators $\abs{F(t)}$ (thick blue) for atom numbers $N = 50,100,200,300,400,500$ (light to dark), $k = 3$, $\phi = 1/\sqrt{N}$, and initial state $e^{- i S_y \pi / 4} e^{- i S_z \pi / 4} \left| \hat{x} \right\rangle$. (b) Unitary (solid) and dissipative (dashed) evolution of $\abs{G(t)}$ (thin yellow) and $\abs{F(t)}$ (thick blue) for $N = 100$.  Dissipative evolution is calculated at $\eta = 100$ and $\delta = 10 \kappa$ from 200 quantum trajectories; error bars are statistical. Horizontal axes show kick number (black) and mean number of photons lost by decay processes in measuring $F$ (blue) and $G$ (yellow) \cite{SM}.}
\label{fig:kt_ip_performance}
\end{figure}

We compare the OTO correlator $F(t)$ with a time-ordered correlator $G(t) = \langle V_t^\dagger V \rangle$, similar to a Loschmidt echo, for the kicked top at several atom numbers $N=2S$ in Fig. \ref{fig:kt_ip_performance}.  We take $V$ and $W$ to be $S_z$ rotations by a small angle $\phi = 1/\sqrt{2S}$, which is chosen so that we can expect to observe a separation of timescales between time-ordered and OTO correlators as $S \rightarrow \infty$ \cite{SM}.  We plot both correlators for an initial coherent state $e^{- i S_y \pi / 4} e^{- i S_z \pi / 4} |S_x = S\rangle$ and kick strength $k=3$, first assuming unitary dynamics.  Even at finite $N$, the OTO correlator (blue) decays on a significantly longer timescale than the time-ordered correlator (yellow).  While the decay time for the time-ordered correlator is roughly independent of $N$, the decay time for the OTO correlator grows as $\log(N)$. This scaling is consistent with a butterfly effect wherein the initial coherent spin state of solid angle $1/N$ exponentially expands on the Bloch sphere.


The measurement of $F$ can be compromised by two forms of dissipation: leakage of photons from the cavity of linewidth $\kappa$; and decay from the atomic excited state of linewidth $\Gamma$.  The fidelities of the controlled phase gate and of the time-reversed Hamiltonian are thus limited by the cooperativity $\eta=4 g^2 / \kappa \Gamma$, where $2g$ is the vacuum Rabi frequency. For an ensemble of $N$ atoms, the maximum achievable controlled phase rotation is of order $\sqrt{\eta/N}$, while observing the onset of chaos in the kicked top requires $\eta \gtrsim (k/2 \ln N)^2$ \cite{SM}.  Thus, dissipation can be kept small at atom numbers $N\lesssim 10^2$ in a state-of-the-art strong coupling cavity with $\eta\sim 10^1 - 10^2$ \cite{Colombe07,Klinder15}, but it cannot be entirely neglected.

To verify that a realistic non-unitary evolution suffices to estimate the OTO correlator, we simulate measurements of $F$ and $G$ in the kicked-top model using quantum trajectories \cite{SM}.  The results of the interferometric protocol are plotted in Fig. \ref{fig:kt_ip_performance}b for a cavity cooperativity $\eta = 100$ (dashed lines) and compared with the unitary case (solid lines). The early-time dissipative evolution is faithful to the unitary evolution, and the difference in timescales between $F$ and $G$ can easily be resolved.  Fully investigating the dissipative effects, by experimental study of longer times and larger atom numbers, may shed new light on chaos and the quantum-to-classical transition in many-body systems.

\textit{Outlook}.  Observing the early-time physics of the OTO correlator in state-of-the-art cavities will allow for probing scrambling in diverse spin models with non-local interactions.  Realizing a model known to have the scrambling properties of a black hole remains a highly nontrivial task.  However, Kitaev has designed one such model, involving random four-fermion interactions \cite{kitaev4f}, that is a close relative of random non-local spin models proposed to study quantum spin glasses \cite{sachdevye} in multimode cavities \cite{multimode,subircavity1}.  Here, periodically modulating external fields or interactions might promote scrambling by melting glass order or by simulating multispin couplings \cite{Rigol14}.

While fast scrambling is a necessary condition for duality to a black hole, whether it is also a sufficient condition is an open question.  If so, then the OTO protocol provides a sharp experimental test for the presence of a black hole by revealing a universal Lyapunov exponent that can be compared with the chaos bound \cite{BHandBF-SS2013,chaosbound}.  A recently proposed cold-atom realization of Kitaev's model \cite{2016arXiv160602454D} could be a platform for validating such a test.


In other physical systems, measurements of the OTO correlator could reveal emergent low-energy bounds on information propagation \cite{2016arXiv160309298R} and distinguish between single-particle \cite{PhysRev.109.1492,billy2008direct,roati2008anderson} and many-body \cite{2006AnPhy.321.1126B,Schreiber842} localization.  Our protocol can be translated directly to trapped-ion simulations of transverse-field Ising models with tunable range \cite{Richerme14,Jurcevic14,bohnet2015quantum}. Neutral Rydberg atoms also allow for engineering local spin models with either sign of interaction \cite{vanBijnen15,Glaetzle15,Zeiher16} and qubit-controlled rotations \cite{rydberggate}.  In optical-lattice implementations of Hubbard models, the sign of interactions can be controlled using Feshbach resonances, the sign of the hopping can be changed by modulating the lattice \cite{shakegeneralproposal,shake4,raman_bloch}, and a controlled phase shift can be imprinted using an impurity atom \cite{Knap12,PhysRevLett.115.135302} or a locally addressed control atom  \cite{bakr2009quantum,weitenberg2011single}.  Alternatively, the distinguishability protocol can be performed by time-of-flight or \textit{in situ} imaging for special initial states, e.g., superfluids or Mott insulators in two dimensions.


\begin{acknowledgments}
M.S.-S. and G.~B. acknowledge support from the NSF, the AFOSR, and the Alfred P. Sloan Foundation.  P.~H. and B.~S. appreciate support from the Simons Foundation and CIFAR. We thank S. Shenker, N. Yao, and E. Demler for enlightening discussions, and we thank E. Davis for feedback on the manuscript.
\end{acknowledgments}

\bibliography{chaos_corr}

\begin{thebibliography}{78}%
\makeatletter
\providecommand \@ifxundefined [1]{%
 \@ifx{#1\undefined}
}%
\providecommand \@ifnum [1]{%
 \ifnum #1\expandafter \@firstoftwo
 \else \expandafter \@secondoftwo
 \fi
}%
\providecommand \@ifx [1]{%
 \ifx #1\expandafter \@firstoftwo
 \else \expandafter \@secondoftwo
 \fi
}%
\providecommand \natexlab [1]{#1}%
\providecommand \enquote  [1]{``#1''}%
\providecommand \bibnamefont  [1]{#1}%
\providecommand \bibfnamefont [1]{#1}%
\providecommand \citenamefont [1]{#1}%
\providecommand \href@noop [0]{\@secondoftwo}%
\providecommand \href [0]{\begingroup \@sanitize@url \@href}%
\providecommand \@href[1]{\@@startlink{#1}\@@href}%
\providecommand \@@href[1]{\endgroup#1\@@endlink}%
\providecommand \@sanitize@url [0]{\catcode `\\12\catcode `\$12\catcode
  `\&12\catcode `\#12\catcode `\^12\catcode `\_12\catcode `\%12\relax}%
\providecommand \@@startlink[1]{}%
\providecommand \@@endlink[0]{}%
\providecommand \url  [0]{\begingroup\@sanitize@url \@url }%
\providecommand \@url [1]{\endgroup\@href {#1}{\urlprefix }}%
\providecommand \urlprefix  [0]{URL }%
\providecommand \Eprint [0]{\href }%
\providecommand \doibase [0]{http://dx.doi.org/}%
\providecommand \selectlanguage [0]{\@gobble}%
\providecommand \bibinfo  [0]{\@secondoftwo}%
\providecommand \bibfield  [0]{\@secondoftwo}%
\providecommand \translation [1]{[#1]}%
\providecommand \BibitemOpen [0]{}%
\providecommand \bibitemStop [0]{}%
\providecommand \bibitemNoStop [0]{.\EOS\space}%
\providecommand \EOS [0]{\spacefactor3000\relax}%
\providecommand \BibitemShut  [1]{\csname bibitem#1\endcsname}%
\let\auto@bib@innerbib\@empty
\bibitem [{\citenamefont {Cheneau}\ \emph {et~al.}(2012)\citenamefont
  {Cheneau}, \citenamefont {Barmettler}, \citenamefont {Poletti}, \citenamefont
  {Endres}, \citenamefont {Schau{\ss}}, \citenamefont {Fukuhara}, \citenamefont
  {Gross}, \citenamefont {Bloch}, \citenamefont {Kollath},\ and\ \citenamefont
  {Kuhr}}]{Cheneau12}%
  \BibitemOpen
  \bibfield  {author} {\bibinfo {author} {\bibfnamefont {M.}~\bibnamefont
  {Cheneau}}, \bibinfo {author} {\bibfnamefont {P.}~\bibnamefont {Barmettler}},
  \bibinfo {author} {\bibfnamefont {D.}~\bibnamefont {Poletti}}, \bibinfo
  {author} {\bibfnamefont {M.}~\bibnamefont {Endres}}, \bibinfo {author}
  {\bibfnamefont {P.}~\bibnamefont {Schau{\ss}}}, \bibinfo {author}
  {\bibfnamefont {T.}~\bibnamefont {Fukuhara}}, \bibinfo {author}
  {\bibfnamefont {C.}~\bibnamefont {Gross}}, \bibinfo {author} {\bibfnamefont
  {I.}~\bibnamefont {Bloch}}, \bibinfo {author} {\bibfnamefont
  {C.}~\bibnamefont {Kollath}}, \ and\ \bibinfo {author} {\bibfnamefont
  {S.}~\bibnamefont {Kuhr}},\ }\href@noop {} {\bibfield  {journal} {\bibinfo
  {journal} {Nature}\ }\textbf {\bibinfo {volume} {481}},\ \bibinfo {pages}
  {484} (\bibinfo {year} {2012})}\BibitemShut {NoStop}%
\bibitem [{\citenamefont {Richerme}\ \emph {et~al.}(2014)\citenamefont
  {Richerme}, \citenamefont {Gong}, \citenamefont {Lee}, \citenamefont {Senko},
  \citenamefont {Smith}, \citenamefont {Foss-Feig}, \citenamefont {Michalakis},
  \citenamefont {Gorshkov},\ and\ \citenamefont {Monroe}}]{Richerme14}%
  \BibitemOpen
  \bibfield  {author} {\bibinfo {author} {\bibfnamefont {P.}~\bibnamefont
  {Richerme}}, \bibinfo {author} {\bibfnamefont {Z.-X.}\ \bibnamefont {Gong}},
  \bibinfo {author} {\bibfnamefont {A.}~\bibnamefont {Lee}}, \bibinfo {author}
  {\bibfnamefont {C.}~\bibnamefont {Senko}}, \bibinfo {author} {\bibfnamefont
  {J.}~\bibnamefont {Smith}}, \bibinfo {author} {\bibfnamefont
  {M.}~\bibnamefont {Foss-Feig}}, \bibinfo {author} {\bibfnamefont
  {S.}~\bibnamefont {Michalakis}}, \bibinfo {author} {\bibfnamefont {A.~V.}\
  \bibnamefont {Gorshkov}}, \ and\ \bibinfo {author} {\bibfnamefont
  {C.}~\bibnamefont {Monroe}},\ }\href@noop {} {\bibfield  {journal} {\bibinfo
  {journal} {Nature}\ }\textbf {\bibinfo {volume} {511}},\ \bibinfo {pages}
  {198} (\bibinfo {year} {2014})}\BibitemShut {NoStop}%
\bibitem [{\citenamefont {Jurcevic}\ \emph {et~al.}(2014)\citenamefont
  {Jurcevic}, \citenamefont {Lanyon}, \citenamefont {Hauke}, \citenamefont
  {Hempel}, \citenamefont {Zoller}, \citenamefont {Blatt},\ and\ \citenamefont
  {Roos}}]{Jurcevic14}%
  \BibitemOpen
  \bibfield  {author} {\bibinfo {author} {\bibfnamefont {P.}~\bibnamefont
  {Jurcevic}}, \bibinfo {author} {\bibfnamefont {B.~P.}\ \bibnamefont
  {Lanyon}}, \bibinfo {author} {\bibfnamefont {P.}~\bibnamefont {Hauke}},
  \bibinfo {author} {\bibfnamefont {C.}~\bibnamefont {Hempel}}, \bibinfo
  {author} {\bibfnamefont {P.}~\bibnamefont {Zoller}}, \bibinfo {author}
  {\bibfnamefont {R.}~\bibnamefont {Blatt}}, \ and\ \bibinfo {author}
  {\bibfnamefont {C.~F.}\ \bibnamefont {Roos}},\ }\href@noop {} {\bibfield
  {journal} {\bibinfo  {journal} {Nature}\ }\textbf {\bibinfo {volume} {511}},\
  \bibinfo {pages} {202} (\bibinfo {year} {2014})}\BibitemShut {NoStop}%
\bibitem [{\citenamefont {Eisert}\ \emph {et~al.}(2015)\citenamefont {Eisert},
  \citenamefont {Friesdorf},\ and\ \citenamefont {Gogolin}}]{Eisert15}%
  \BibitemOpen
  \bibfield  {author} {\bibinfo {author} {\bibfnamefont {J.}~\bibnamefont
  {Eisert}}, \bibinfo {author} {\bibfnamefont {M.}~\bibnamefont {Friesdorf}}, \
  and\ \bibinfo {author} {\bibfnamefont {C.}~\bibnamefont {Gogolin}},\ }\href
  {http://dx.doi.org/10.1038/nphys3215} {\bibfield  {journal} {\bibinfo
  {journal} {Nat Phys}\ }\textbf {\bibinfo {volume} {11}},\ \bibinfo {pages}
  {124} (\bibinfo {year} {2015})}\BibitemShut {NoStop}%
\bibitem [{\citenamefont {Lieb}\ and\ \citenamefont
  {Robinson}(1972)}]{lieb72finite}%
  \BibitemOpen
  \bibfield  {author} {\bibinfo {author} {\bibfnamefont {E.}~\bibnamefont
  {Lieb}}\ and\ \bibinfo {author} {\bibfnamefont {D.}~\bibnamefont
  {Robinson}},\ }\href@noop {} {\bibfield  {journal} {\bibinfo  {journal}
  {Communications in Mathematical Physics}\ }\textbf {\bibinfo {volume} {28}},\
  \bibinfo {pages} {251} (\bibinfo {year} {1972})}\BibitemShut {NoStop}%
\bibitem [{\citenamefont {{Shenker}}\ and\ \citenamefont
  {{Stanford}}(2014)}]{BHandBF-SS2013}%
  \BibitemOpen
  \bibfield  {author} {\bibinfo {author} {\bibfnamefont {S.~H.}\ \bibnamefont
  {{Shenker}}}\ and\ \bibinfo {author} {\bibfnamefont {D.}~\bibnamefont
  {{Stanford}}},\ }\href {\doibase 10.1007/JHEP03(2014)067} {\bibfield
  {journal} {\bibinfo  {journal} {Journal of High Energy Physics}\ }\textbf
  {\bibinfo {volume} {3}},\ \bibinfo {eid} {67} (\bibinfo {year} {2014})},\
  \Eprint {http://arxiv.org/abs/1306.0622} {arXiv:1306.0622 [hep-th]}
  \BibitemShut {NoStop}%
\bibitem [{\citenamefont {Bekenstein}(1973)}]{Bekenstein:1973ur}%
  \BibitemOpen
  \bibfield  {author} {\bibinfo {author} {\bibfnamefont {J.~D.}\ \bibnamefont
  {Bekenstein}},\ }\href {\doibase 10.1103/PhysRevD.7.2333} {\bibfield
  {journal} {\bibinfo  {journal} {Phys. Rev.}\ }\textbf {\bibinfo {volume}
  {D7}},\ \bibinfo {pages} {2333} (\bibinfo {year} {1973})}\BibitemShut
  {NoStop}%
\bibitem [{\citenamefont {{Hayden}}\ and\ \citenamefont
  {{Preskill}}(2007)}]{haydenpreskill}%
  \BibitemOpen
  \bibfield  {author} {\bibinfo {author} {\bibfnamefont {P.}~\bibnamefont
  {{Hayden}}}\ and\ \bibinfo {author} {\bibfnamefont {J.}~\bibnamefont
  {{Preskill}}},\ }\href {\doibase 10.1088/1126-6708/2007/09/120} {\bibfield
  {journal} {\bibinfo  {journal} {Journal of High Energy Physics}\ }\textbf
  {\bibinfo {volume} {9}},\ \bibinfo {eid} {120} (\bibinfo {year} {2007})},\
  \Eprint {http://arxiv.org/abs/0708.4025} {arXiv:0708.4025 [hep-th]}
  \BibitemShut {NoStop}%
\bibitem [{\citenamefont {{Sekino}}\ and\ \citenamefont
  {{Susskind}}(2008)}]{fastscramble}%
  \BibitemOpen
  \bibfield  {author} {\bibinfo {author} {\bibfnamefont {Y.}~\bibnamefont
  {{Sekino}}}\ and\ \bibinfo {author} {\bibfnamefont {L.}~\bibnamefont
  {{Susskind}}},\ }\href {\doibase 10.1088/1126-6708/2008/10/065} {\bibfield
  {journal} {\bibinfo  {journal} {Journal of High Energy Physics}\ }\textbf
  {\bibinfo {volume} {10}},\ \bibinfo {eid} {065} (\bibinfo {year} {2008})},\
  \Eprint {http://arxiv.org/abs/0808.2096} {arXiv:0808.2096 [hep-th]}
  \BibitemShut {NoStop}%
\bibitem [{\citenamefont {{Maldacena}}\ \emph {et~al.}(2015)\citenamefont
  {{Maldacena}}, \citenamefont {{Shenker}},\ and\ \citenamefont
  {{Stanford}}}]{chaosbound}%
  \BibitemOpen
  \bibfield  {author} {\bibinfo {author} {\bibfnamefont {J.}~\bibnamefont
  {{Maldacena}}}, \bibinfo {author} {\bibfnamefont {S.~H.}\ \bibnamefont
  {{Shenker}}}, \ and\ \bibinfo {author} {\bibfnamefont {D.}~\bibnamefont
  {{Stanford}}},\ }\href@noop {} {\bibfield  {journal} {\bibinfo  {journal}
  {ArXiv e-prints}\ } (\bibinfo {year} {2015})},\ \Eprint
  {http://arxiv.org/abs/1503.01409} {arXiv:1503.01409 [hep-th]} \BibitemShut
  {NoStop}%
\bibitem [{\citenamefont {Hooft}(1993)}]{hooft1993dimensional}%
  \BibitemOpen
  \bibfield  {author} {\bibinfo {author} {\bibfnamefont {G.}~\bibnamefont
  {Hooft}},\ }\href@noop {} {\bibfield  {journal} {\bibinfo  {journal} {arXiv
  preprint gr-qc/9310026}\ } (\bibinfo {year} {1993})}\BibitemShut {NoStop}%
\bibitem [{\citenamefont {Susskind}(1995)}]{susskind1995world}%
  \BibitemOpen
  \bibfield  {author} {\bibinfo {author} {\bibfnamefont {L.}~\bibnamefont
  {Susskind}},\ }\href@noop {} {\bibfield  {journal} {\bibinfo  {journal}
  {Journal of Mathematical Physics}\ }\textbf {\bibinfo {volume} {36}},\
  \bibinfo {pages} {6377} (\bibinfo {year} {1995})}\BibitemShut {NoStop}%
\bibitem [{\citenamefont {{Maldacena}}(1999)}]{adscft}%
  \BibitemOpen
  \bibfield  {author} {\bibinfo {author} {\bibfnamefont {J.}~\bibnamefont
  {{Maldacena}}},\ }\href {\doibase 10.1023/A:1026654312961} {\bibfield
  {journal} {\bibinfo  {journal} {International Journal of Theoretical
  Physics}\ }\textbf {\bibinfo {volume} {38}},\ \bibinfo {pages} {1113}
  (\bibinfo {year} {1999})},\ \Eprint {http://arxiv.org/abs/hep-th/9711200}
  {hep-th/9711200} \BibitemShut {NoStop}%
\bibitem [{\citenamefont {{Banks}}\ \emph {et~al.}(1997)\citenamefont
  {{Banks}}, \citenamefont {{Fischler}}, \citenamefont {{Shenker}},\ and\
  \citenamefont {{Susskind}}}]{bfss}%
  \BibitemOpen
  \bibfield  {author} {\bibinfo {author} {\bibfnamefont {T.}~\bibnamefont
  {{Banks}}}, \bibinfo {author} {\bibfnamefont {W.}~\bibnamefont {{Fischler}}},
  \bibinfo {author} {\bibfnamefont {S.~H.}\ \bibnamefont {{Shenker}}}, \ and\
  \bibinfo {author} {\bibfnamefont {L.}~\bibnamefont {{Susskind}}},\ }\href
  {\doibase 10.1103/PhysRevD.55.5112} {\bibfield  {journal} {\bibinfo
  {journal} {\prd}\ }\textbf {\bibinfo {volume} {55}},\ \bibinfo {pages} {5112}
  (\bibinfo {year} {1997})},\ \Eprint {http://arxiv.org/abs/hep-th/9610043}
  {hep-th/9610043} \BibitemShut {NoStop}%
\bibitem [{\citenamefont {{Aharony}}\ \emph {et~al.}(2000)\citenamefont
  {{Aharony}}, \citenamefont {{Gubser}}, \citenamefont {{Maldacena}},
  \citenamefont {{Ooguri}},\ and\ \citenamefont {{Oz}}}]{magoo}%
  \BibitemOpen
  \bibfield  {author} {\bibinfo {author} {\bibfnamefont {O.}~\bibnamefont
  {{Aharony}}}, \bibinfo {author} {\bibfnamefont {S.~S.}\ \bibnamefont
  {{Gubser}}}, \bibinfo {author} {\bibfnamefont {J.}~\bibnamefont
  {{Maldacena}}}, \bibinfo {author} {\bibfnamefont {H.}~\bibnamefont
  {{Ooguri}}}, \ and\ \bibinfo {author} {\bibfnamefont {Y.}~\bibnamefont
  {{Oz}}},\ }\href {\doibase 10.1016/S0370-1573(99)00083-6} {\bibfield
  {journal} {\bibinfo  {journal} {Phys. Rep.}\ }\textbf {\bibinfo {volume}
  {323}},\ \bibinfo {pages} {183} (\bibinfo {year} {2000})},\ \Eprint
  {http://arxiv.org/abs/hep-th/9905111} {hep-th/9905111} \BibitemShut {NoStop}%
\bibitem [{\citenamefont {{Zohar}}\ \emph {et~al.}(2016)\citenamefont
  {{Zohar}}, \citenamefont {{Cirac}},\ and\ \citenamefont
  {{Reznik}}}]{cold_atoms_gauge}%
  \BibitemOpen
  \bibfield  {author} {\bibinfo {author} {\bibfnamefont {E.}~\bibnamefont
  {{Zohar}}}, \bibinfo {author} {\bibfnamefont {J.~I.}\ \bibnamefont
  {{Cirac}}}, \ and\ \bibinfo {author} {\bibfnamefont {B.}~\bibnamefont
  {{Reznik}}},\ }\href {\doibase 10.1088/0034-4885/79/1/014401} {\bibfield
  {journal} {\bibinfo  {journal} {Reports on Progress in Physics}\ }\textbf
  {\bibinfo {volume} {79}},\ \bibinfo {eid} {014401} (\bibinfo {year}
  {2016})},\ \Eprint {http://arxiv.org/abs/1503.02312} {arXiv:1503.02312
  [quant-ph]} \BibitemShut {NoStop}%
\bibitem [{\citenamefont {Hosur}\ \emph {et~al.}(2016)\citenamefont {Hosur},
  \citenamefont {Qi}, \citenamefont {Roberts},\ and\ \citenamefont
  {Yoshida}}]{Hosur2016}%
  \BibitemOpen
  \bibfield  {author} {\bibinfo {author} {\bibfnamefont {P.}~\bibnamefont
  {Hosur}}, \bibinfo {author} {\bibfnamefont {X.-L.}\ \bibnamefont {Qi}},
  \bibinfo {author} {\bibfnamefont {D.~A.}\ \bibnamefont {Roberts}}, \ and\
  \bibinfo {author} {\bibfnamefont {B.}~\bibnamefont {Yoshida}},\ }\href@noop
  {} {\bibfield  {journal} {\bibinfo  {journal} {Journal of High Energy
  Physics}\ ,\ \bibinfo {pages} {1}} (\bibinfo {year} {2016})}\BibitemShut
  {NoStop}%
\bibitem [{\citenamefont {S{\o}rensen}\ and\ \citenamefont
  {M{\o}lmer}(2002)}]{sorensen02}%
  \BibitemOpen
  \bibfield  {author} {\bibinfo {author} {\bibfnamefont {A.~S.}\ \bibnamefont
  {S{\o}rensen}}\ and\ \bibinfo {author} {\bibfnamefont {K.}~\bibnamefont
  {M{\o}lmer}},\ }\href@noop {} {\bibfield  {journal} {\bibinfo  {journal}
  {Physical Review A}\ }\textbf {\bibinfo {volume} {66}},\ \bibinfo {pages}
  {022314} (\bibinfo {year} {2002})}\BibitemShut {NoStop}%
\bibitem [{\citenamefont {Leroux}\ \emph {et~al.}(2010)\citenamefont {Leroux},
  \citenamefont {Schleier-Smith},\ and\ \citenamefont
  {Vuleti{\'c}}}]{leroux2010implementation}%
  \BibitemOpen
  \bibfield  {author} {\bibinfo {author} {\bibfnamefont {I.~D.}\ \bibnamefont
  {Leroux}}, \bibinfo {author} {\bibfnamefont {M.~H.}\ \bibnamefont
  {Schleier-Smith}}, \ and\ \bibinfo {author} {\bibfnamefont {V.}~\bibnamefont
  {Vuleti{\'c}}},\ }\href@noop {} {\bibfield  {journal} {\bibinfo  {journal}
  {Physical Review Letters}\ }\textbf {\bibinfo {volume} {104}},\ \bibinfo
  {pages} {073602} (\bibinfo {year} {2010})}\BibitemShut {NoStop}%
\bibitem [{\citenamefont {Gopalakrishnan}\ \emph {et~al.}(2011)\citenamefont
  {Gopalakrishnan}, \citenamefont {Lev},\ and\ \citenamefont
  {Goldbart}}]{multimode}%
  \BibitemOpen
  \bibfield  {author} {\bibinfo {author} {\bibfnamefont {S.}~\bibnamefont
  {Gopalakrishnan}}, \bibinfo {author} {\bibfnamefont {B.~L.}\ \bibnamefont
  {Lev}}, \ and\ \bibinfo {author} {\bibfnamefont {P.~M.}\ \bibnamefont
  {Goldbart}},\ }\href {\doibase 10.1103/PhysRevLett.107.277201} {\bibfield
  {journal} {\bibinfo  {journal} {Phys. Rev. Lett.}\ }\textbf {\bibinfo
  {volume} {107}},\ \bibinfo {pages} {277201} (\bibinfo {year}
  {2011})}\BibitemShut {NoStop}%
\bibitem [{\citenamefont {{Strack}}\ and\ \citenamefont
  {{Sachdev}}(2011)}]{subircavity1}%
  \BibitemOpen
  \bibfield  {author} {\bibinfo {author} {\bibfnamefont {P.}~\bibnamefont
  {{Strack}}}\ and\ \bibinfo {author} {\bibfnamefont {S.}~\bibnamefont
  {{Sachdev}}},\ }\href {\doibase 10.1103/PhysRevLett.107.277202} {\bibfield
  {journal} {\bibinfo  {journal} {Physical Review Letters}\ }\textbf {\bibinfo
  {volume} {107}},\ \bibinfo {eid} {277202} (\bibinfo {year} {2011})},\ \Eprint
  {http://arxiv.org/abs/1109.2119} {arXiv:1109.2119 [cond-mat.quant-gas]}
  \BibitemShut {NoStop}%
\bibitem [{\citenamefont {Hosten}\ \emph {et~al.}(2016)\citenamefont {Hosten},
  \citenamefont {Krishnakumar}, \citenamefont {Engelsen},\ and\ \citenamefont
  {Kasevich}}]{hosten2016quantum}%
  \BibitemOpen
  \bibfield  {author} {\bibinfo {author} {\bibfnamefont {O.}~\bibnamefont
  {Hosten}}, \bibinfo {author} {\bibfnamefont {R.}~\bibnamefont
  {Krishnakumar}}, \bibinfo {author} {\bibfnamefont {N.~J.}\ \bibnamefont
  {Engelsen}}, \ and\ \bibinfo {author} {\bibfnamefont {M.~A.}\ \bibnamefont
  {Kasevich}},\ }\href@noop {} {\bibfield  {journal} {\bibinfo  {journal}
  {Science}\ }\textbf {\bibinfo {volume} {352}},\ \bibinfo {pages} {1552}
  (\bibinfo {year} {2016})}\BibitemShut {NoStop}%
\bibitem [{\citenamefont {Lashkari}\ \emph {et~al.}(2013)\citenamefont
  {Lashkari}, \citenamefont {Stanford}, \citenamefont {Hastings}, \citenamefont
  {Osborne},\ and\ \citenamefont {Hayden}}]{Lashkari13}%
  \BibitemOpen
  \bibfield  {author} {\bibinfo {author} {\bibfnamefont {N.}~\bibnamefont
  {Lashkari}}, \bibinfo {author} {\bibfnamefont {D.}~\bibnamefont {Stanford}},
  \bibinfo {author} {\bibfnamefont {M.}~\bibnamefont {Hastings}}, \bibinfo
  {author} {\bibfnamefont {T.}~\bibnamefont {Osborne}}, \ and\ \bibinfo
  {author} {\bibfnamefont {P.}~\bibnamefont {Hayden}},\ }\href@noop {}
  {\bibfield  {journal} {\bibinfo  {journal} {Journal of High Energy Physics}\
  }\textbf {\bibinfo {volume} {2013}},\ \bibinfo {pages} {1} (\bibinfo {year}
  {2013})}\BibitemShut {NoStop}%
\bibitem [{\citenamefont {Brown}\ \emph {et~al.}(2016)\citenamefont {Brown},
  \citenamefont {Roberts}, \citenamefont {Susskind}, \citenamefont {Swingle},\
  and\ \citenamefont {Zhao}}]{Brown2016}%
  \BibitemOpen
  \bibfield  {author} {\bibinfo {author} {\bibfnamefont {A.~R.}\ \bibnamefont
  {Brown}}, \bibinfo {author} {\bibfnamefont {D.~A.}\ \bibnamefont {Roberts}},
  \bibinfo {author} {\bibfnamefont {L.}~\bibnamefont {Susskind}}, \bibinfo
  {author} {\bibfnamefont {B.}~\bibnamefont {Swingle}}, \ and\ \bibinfo
  {author} {\bibfnamefont {Y.}~\bibnamefont {Zhao}},\ }\href {\doibase
  10.1103/PhysRevLett.116.191301} {\bibfield  {journal} {\bibinfo  {journal}
  {Phys. Rev. Lett.}\ }\textbf {\bibinfo {volume} {116}},\ \bibinfo {pages}
  {191301} (\bibinfo {year} {2016})}\BibitemShut {NoStop}%
\bibitem [{\citenamefont {{Prosen}}(2014)}]{2014PhRvE..89a2142P}%
  \BibitemOpen
  \bibfield  {author} {\bibinfo {author} {\bibfnamefont {T.}~\bibnamefont
  {{Prosen}}},\ }\href {\doibase 10.1103/PhysRevE.89.012142} {\bibfield
  {journal} {\bibinfo  {journal} {\pre}\ }\textbf {\bibinfo {volume} {89}},\
  \bibinfo {eid} {012142} (\bibinfo {year} {2014})},\ \Eprint
  {http://arxiv.org/abs/1310.8629} {arXiv:1310.8629 [cond-mat.stat-mech]}
  \BibitemShut {NoStop}%
\bibitem [{\citenamefont {{Roberts}}\ and\ \citenamefont
  {{Swingle}}(2016)}]{2016arXiv160309298R}%
  \BibitemOpen
  \bibfield  {author} {\bibinfo {author} {\bibfnamefont {D.~A.}\ \bibnamefont
  {{Roberts}}}\ and\ \bibinfo {author} {\bibfnamefont {B.}~\bibnamefont
  {{Swingle}}},\ }\href@noop {} {\bibfield  {journal} {\bibinfo  {journal}
  {ArXiv e-prints}\ } (\bibinfo {year} {2016})},\ \Eprint
  {http://arxiv.org/abs/1603.09298} {arXiv:1603.09298 [hep-th]} \BibitemShut
  {NoStop}%
\bibitem [{\citenamefont {{Blake}}(2016)}]{2016arXiv160308510B}%
  \BibitemOpen
  \bibfield  {author} {\bibinfo {author} {\bibfnamefont {M.}~\bibnamefont
  {{Blake}}},\ }\href@noop {} {\bibfield  {journal} {\bibinfo  {journal} {ArXiv
  e-prints}\ } (\bibinfo {year} {2016})},\ \Eprint
  {http://arxiv.org/abs/1603.08510} {arXiv:1603.08510 [hep-th]} \BibitemShut
  {NoStop}%
\bibitem [{\citenamefont {{Kitaev}}(2014)}]{kitaevbhchaos}%
  \BibitemOpen
  \bibfield  {author} {\bibinfo {author} {\bibfnamefont {A.}~\bibnamefont
  {{Kitaev}}},\ }\href@noop {} {\bibfield  {journal} {\bibinfo  {journal} {talk
  at Fundamental Physics Prize Symposium Nov. 10, 2014}\ } (\bibinfo {year}
  {2014})}\BibitemShut {NoStop}%
\bibitem [{\citenamefont {Shenker}\ and\ \citenamefont
  {Stanford}(2014)}]{shenker2014multiple}%
  \BibitemOpen
  \bibfield  {author} {\bibinfo {author} {\bibfnamefont {S.~H.}\ \bibnamefont
  {Shenker}}\ and\ \bibinfo {author} {\bibfnamefont {D.}~\bibnamefont
  {Stanford}},\ }\href@noop {} {\bibfield  {journal} {\bibinfo  {journal}
  {Journal of High Energy Physics}\ }\textbf {\bibinfo {volume} {2014}},\
  \bibinfo {pages} {1} (\bibinfo {year} {2014})}\BibitemShut {NoStop}%
\bibitem [{\citenamefont {Roberts}\ \emph {et~al.}(2015)\citenamefont
  {Roberts}, \citenamefont {Stanford},\ and\ \citenamefont
  {Susskind}}]{roberts2015localized}%
  \BibitemOpen
  \bibfield  {author} {\bibinfo {author} {\bibfnamefont {D.~A.}\ \bibnamefont
  {Roberts}}, \bibinfo {author} {\bibfnamefont {D.}~\bibnamefont {Stanford}}, \
  and\ \bibinfo {author} {\bibfnamefont {L.}~\bibnamefont {Susskind}},\
  }\href@noop {} {\bibfield  {journal} {\bibinfo  {journal} {Journal of High
  Energy Physics}\ }\textbf {\bibinfo {volume} {2015}},\ \bibinfo {pages} {1}
  (\bibinfo {year} {2015})}\BibitemShut {NoStop}%
\bibitem [{\citenamefont {{Goussev}}\ \emph {et~al.}(2012)\citenamefont
  {{Goussev}}, \citenamefont {{Jalabert}}, \citenamefont {{Pastawski}},\ and\
  \citenamefont {{Wisniacki}}}]{EchoReview}%
  \BibitemOpen
  \bibfield  {author} {\bibinfo {author} {\bibfnamefont {A.}~\bibnamefont
  {{Goussev}}}, \bibinfo {author} {\bibfnamefont {R.~A.}\ \bibnamefont
  {{Jalabert}}}, \bibinfo {author} {\bibfnamefont {H.~M.}\ \bibnamefont
  {{Pastawski}}}, \ and\ \bibinfo {author} {\bibfnamefont {D.}~\bibnamefont
  {{Wisniacki}}},\ }\href@noop {} {\bibfield  {journal} {\bibinfo  {journal}
  {ArXiv e-prints}\ } (\bibinfo {year} {2012})},\ \Eprint
  {http://arxiv.org/abs/1206.6348} {arXiv:1206.6348 [nlin.CD]} \BibitemShut
  {NoStop}%
\bibitem [{\citenamefont {Hahn}(1950)}]{HahnEcho}%
  \BibitemOpen
  \bibfield  {author} {\bibinfo {author} {\bibfnamefont {E.~L.}\ \bibnamefont
  {Hahn}},\ }\href {\doibase 10.1103/PhysRev.80.580} {\bibfield  {journal}
  {\bibinfo  {journal} {Phys. Rev.}\ }\textbf {\bibinfo {volume} {80}},\
  \bibinfo {pages} {580} (\bibinfo {year} {1950})}\BibitemShut {NoStop}%
\bibitem [{\citenamefont {Rhim}\ \emph {et~al.}(1971)\citenamefont {Rhim},
  \citenamefont {Pines},\ and\ \citenamefont {Waugh}}]{MagicEcho}%
  \BibitemOpen
  \bibfield  {author} {\bibinfo {author} {\bibfnamefont {W.-K.}\ \bibnamefont
  {Rhim}}, \bibinfo {author} {\bibfnamefont {A.}~\bibnamefont {Pines}}, \ and\
  \bibinfo {author} {\bibfnamefont {J.~S.}\ \bibnamefont {Waugh}},\ }\href
  {\doibase 10.1103/PhysRevB.3.684} {\bibfield  {journal} {\bibinfo  {journal}
  {Phys. Rev. B}\ }\textbf {\bibinfo {volume} {3}},\ \bibinfo {pages} {684}
  (\bibinfo {year} {1971})}\BibitemShut {NoStop}%
\bibitem [{\citenamefont {Zhang}\ \emph {et~al.}(1992)\citenamefont {Zhang},
  \citenamefont {Meier},\ and\ \citenamefont {Ernst}}]{PolarizationEcho}%
  \BibitemOpen
  \bibfield  {author} {\bibinfo {author} {\bibfnamefont {S.}~\bibnamefont
  {Zhang}}, \bibinfo {author} {\bibfnamefont {B.~H.}\ \bibnamefont {Meier}}, \
  and\ \bibinfo {author} {\bibfnamefont {R.~R.}\ \bibnamefont {Ernst}},\ }\href
  {\doibase 10.1103/PhysRevLett.69.2149} {\bibfield  {journal} {\bibinfo
  {journal} {Phys. Rev. Lett.}\ }\textbf {\bibinfo {volume} {69}},\ \bibinfo
  {pages} {2149} (\bibinfo {year} {1992})}\BibitemShut {NoStop}%
\bibitem [{\citenamefont {Andersen}\ \emph {et~al.}(2003)\citenamefont
  {Andersen}, \citenamefont {Kaplan},\ and\ \citenamefont
  {Davidson}}]{AtomEcho}%
  \BibitemOpen
  \bibfield  {author} {\bibinfo {author} {\bibfnamefont {M.~F.}\ \bibnamefont
  {Andersen}}, \bibinfo {author} {\bibfnamefont {A.}~\bibnamefont {Kaplan}}, \
  and\ \bibinfo {author} {\bibfnamefont {N.}~\bibnamefont {Davidson}},\ }\href
  {\doibase 10.1103/PhysRevLett.90.023001} {\bibfield  {journal} {\bibinfo
  {journal} {Phys. Rev. Lett.}\ }\textbf {\bibinfo {volume} {90}},\ \bibinfo
  {pages} {023001} (\bibinfo {year} {2003})}\BibitemShut {NoStop}%
\bibitem [{\citenamefont {Gorin}\ \emph {et~al.}(2006)\citenamefont {Gorin},
  \citenamefont {Prosen}, \citenamefont {Seligman},\ and\ \citenamefont
  {{\v{Z}}nidari{\v{c}}}}]{gorin2006dynamics}%
  \BibitemOpen
  \bibfield  {author} {\bibinfo {author} {\bibfnamefont {T.}~\bibnamefont
  {Gorin}}, \bibinfo {author} {\bibfnamefont {T.}~\bibnamefont {Prosen}},
  \bibinfo {author} {\bibfnamefont {T.~H.}\ \bibnamefont {Seligman}}, \ and\
  \bibinfo {author} {\bibfnamefont {M.}~\bibnamefont {{\v{Z}}nidari{\v{c}}}},\
  }\href@noop {} {\bibfield  {journal} {\bibinfo  {journal} {Physics Reports}\
  }\textbf {\bibinfo {volume} {435}},\ \bibinfo {pages} {33} (\bibinfo {year}
  {2006})}\BibitemShut {NoStop}%
\bibitem [{\citenamefont {Zurek}\ and\ \citenamefont
  {Paz}(1994)}]{DecoherenceChaosSecondLaw}%
  \BibitemOpen
  \bibfield  {author} {\bibinfo {author} {\bibfnamefont {W.~H.}\ \bibnamefont
  {Zurek}}\ and\ \bibinfo {author} {\bibfnamefont {J.~P.}\ \bibnamefont
  {Paz}},\ }\href {\doibase 10.1103/PhysRevLett.72.2508} {\bibfield  {journal}
  {\bibinfo  {journal} {Phys. Rev. Lett.}\ }\textbf {\bibinfo {volume} {72}},\
  \bibinfo {pages} {2508} (\bibinfo {year} {1994})}\BibitemShut {NoStop}%
\bibitem [{\citenamefont {Jalabert}\ and\ \citenamefont
  {Pastawski}(2001)}]{DecoherenceJalabert}%
  \BibitemOpen
  \bibfield  {author} {\bibinfo {author} {\bibfnamefont {R.~A.}\ \bibnamefont
  {Jalabert}}\ and\ \bibinfo {author} {\bibfnamefont {H.~M.}\ \bibnamefont
  {Pastawski}},\ }\href {\doibase 10.1103/PhysRevLett.86.2490} {\bibfield
  {journal} {\bibinfo  {journal} {Phys. Rev. Lett.}\ }\textbf {\bibinfo
  {volume} {86}},\ \bibinfo {pages} {2490} (\bibinfo {year}
  {2001})}\BibitemShut {NoStop}%
\bibitem [{\citenamefont {Cucchietti}\ \emph {et~al.}(2003)\citenamefont
  {Cucchietti}, \citenamefont {Dalvit}, \citenamefont {Paz},\ and\
  \citenamefont {Zurek}}]{DecoherenceEchoZurek}%
  \BibitemOpen
  \bibfield  {author} {\bibinfo {author} {\bibfnamefont {F.~M.}\ \bibnamefont
  {Cucchietti}}, \bibinfo {author} {\bibfnamefont {D.~A.~R.}\ \bibnamefont
  {Dalvit}}, \bibinfo {author} {\bibfnamefont {J.~P.}\ \bibnamefont {Paz}}, \
  and\ \bibinfo {author} {\bibfnamefont {W.~H.}\ \bibnamefont {Zurek}},\ }\href
  {\doibase 10.1103/PhysRevLett.91.210403} {\bibfield  {journal} {\bibinfo
  {journal} {Phys. Rev. Lett.}\ }\textbf {\bibinfo {volume} {91}},\ \bibinfo
  {pages} {210403} (\bibinfo {year} {2003})}\BibitemShut {NoStop}%
\bibitem [{\citenamefont {{Larkin}}\ and\ \citenamefont
  {{Ovchinnikov}}(1969)}]{quasiclassicalsc}%
  \BibitemOpen
  \bibfield  {author} {\bibinfo {author} {\bibfnamefont {A.~I.}\ \bibnamefont
  {{Larkin}}}\ and\ \bibinfo {author} {\bibfnamefont {Y.~N.}\ \bibnamefont
  {{Ovchinnikov}}},\ }\href@noop {} {\bibfield  {journal} {\bibinfo  {journal}
  {Soviet Journal of Experimental and Theoretical Physics}\ }\textbf {\bibinfo
  {volume} {28}},\ \bibinfo {pages} {1200} (\bibinfo {year}
  {1969})}\BibitemShut {NoStop}%
\bibitem [{foo()}]{foot1}%
  \BibitemOpen
  \href@noop {} {\ }\bibinfo {note} {For black holes and some semi-classical
  models there is an additional subtlety: the coefficient of
  $e^{t/\tau}/\mathbb{S}$ in $1-F$ is purely imaginary, so the growth of the
  commutator $\langle \abs{[W_t,V]}^2\rangle$ begins at order
  $(e^{t/\tau}/\mathbb{S})^2$. This does not effect the scrambling
  time.}\BibitemShut {Stop}%
\bibitem [{\citenamefont {Knap}\ \emph {et~al.}(2012)\citenamefont {Knap},
  \citenamefont {Shashi}, \citenamefont {Nishida}, \citenamefont {Imambekov},
  \citenamefont {Abanin},\ and\ \citenamefont {Demler}}]{Knap12}%
  \BibitemOpen
  \bibfield  {author} {\bibinfo {author} {\bibfnamefont {M.}~\bibnamefont
  {Knap}}, \bibinfo {author} {\bibfnamefont {A.}~\bibnamefont {Shashi}},
  \bibinfo {author} {\bibfnamefont {Y.}~\bibnamefont {Nishida}}, \bibinfo
  {author} {\bibfnamefont {A.}~\bibnamefont {Imambekov}}, \bibinfo {author}
  {\bibfnamefont {D.~A.}\ \bibnamefont {Abanin}}, \ and\ \bibinfo {author}
  {\bibfnamefont {E.}~\bibnamefont {Demler}},\ }\href@noop {} {\bibfield
  {journal} {\bibinfo  {journal} {Physical Review X}\ }\textbf {\bibinfo
  {volume} {2}},\ \bibinfo {pages} {041020} (\bibinfo {year}
  {2012})}\BibitemShut {NoStop}%
\bibitem [{\citenamefont {Cetina}\ \emph {et~al.}(2015)\citenamefont {Cetina},
  \citenamefont {Jag}, \citenamefont {Lous}, \citenamefont {Walraven},
  \citenamefont {Grimm}, \citenamefont {Christensen},\ and\ \citenamefont
  {Bruun}}]{PhysRevLett.115.135302}%
  \BibitemOpen
  \bibfield  {author} {\bibinfo {author} {\bibfnamefont {M.}~\bibnamefont
  {Cetina}}, \bibinfo {author} {\bibfnamefont {M.}~\bibnamefont {Jag}},
  \bibinfo {author} {\bibfnamefont {R.~S.}\ \bibnamefont {Lous}}, \bibinfo
  {author} {\bibfnamefont {J.~T.~M.}\ \bibnamefont {Walraven}}, \bibinfo
  {author} {\bibfnamefont {R.}~\bibnamefont {Grimm}}, \bibinfo {author}
  {\bibfnamefont {R.~S.}\ \bibnamefont {Christensen}}, \ and\ \bibinfo {author}
  {\bibfnamefont {G.~M.}\ \bibnamefont {Bruun}},\ }\href {\doibase
  10.1103/PhysRevLett.115.135302} {\bibfield  {journal} {\bibinfo  {journal}
  {Phys. Rev. Lett.}\ }\textbf {\bibinfo {volume} {115}},\ \bibinfo {pages}
  {135302} (\bibinfo {year} {2015})}\BibitemShut {NoStop}%
\bibitem [{\citenamefont {Pedernales}\ \emph {et~al.}(2014)\citenamefont
  {Pedernales}, \citenamefont {Di~Candia}, \citenamefont {Egusquiza},
  \citenamefont {Casanova},\ and\ \citenamefont
  {Solano}}]{PhysRevLett.113.020505}%
  \BibitemOpen
  \bibfield  {author} {\bibinfo {author} {\bibfnamefont {J.~S.}\ \bibnamefont
  {Pedernales}}, \bibinfo {author} {\bibfnamefont {R.}~\bibnamefont
  {Di~Candia}}, \bibinfo {author} {\bibfnamefont {I.~L.}\ \bibnamefont
  {Egusquiza}}, \bibinfo {author} {\bibfnamefont {J.}~\bibnamefont {Casanova}},
  \ and\ \bibinfo {author} {\bibfnamefont {E.}~\bibnamefont {Solano}},\ }\href
  {\doibase 10.1103/PhysRevLett.113.020505} {\bibfield  {journal} {\bibinfo
  {journal} {Phys. Rev. Lett.}\ }\textbf {\bibinfo {volume} {113}},\ \bibinfo
  {pages} {020505} (\bibinfo {year} {2014})}\BibitemShut {NoStop}%
\bibitem [{\citenamefont {{Pedernales}}\ \emph {et~al.}(2014)\citenamefont
  {{Pedernales}}, \citenamefont {{Di Candia}}, \citenamefont {{Egusquiza}},
  \citenamefont {{Casanova}},\ and\ \citenamefont
  {{Solano}}}]{2014PhRvL.113b0505P}%
  \BibitemOpen
  \bibfield  {author} {\bibinfo {author} {\bibfnamefont {J.~S.}\ \bibnamefont
  {{Pedernales}}}, \bibinfo {author} {\bibfnamefont {R.}~\bibnamefont {{Di
  Candia}}}, \bibinfo {author} {\bibfnamefont {I.~L.}\ \bibnamefont
  {{Egusquiza}}}, \bibinfo {author} {\bibfnamefont {J.}~\bibnamefont
  {{Casanova}}}, \ and\ \bibinfo {author} {\bibfnamefont {E.}~\bibnamefont
  {{Solano}}},\ }\href {\doibase 10.1103/PhysRevLett.113.020505} {\bibfield
  {journal} {\bibinfo  {journal} {Physical Review Letters}\ }\textbf {\bibinfo
  {volume} {113}},\ \bibinfo {eid} {020505} (\bibinfo {year} {2014})},\ \Eprint
  {http://arxiv.org/abs/1401.2430} {arXiv:1401.2430 [quant-ph]} \BibitemShut
  {NoStop}%
\bibitem [{\citenamefont {Abanin}\ and\ \citenamefont
  {Demler}(2012)}]{abanin2012measuring}%
  \BibitemOpen
  \bibfield  {author} {\bibinfo {author} {\bibfnamefont {D.~A.}\ \bibnamefont
  {Abanin}}\ and\ \bibinfo {author} {\bibfnamefont {E.}~\bibnamefont
  {Demler}},\ }\href@noop {} {\bibfield  {journal} {\bibinfo  {journal}
  {Physical review letters}\ }\textbf {\bibinfo {volume} {109}},\ \bibinfo
  {pages} {020504} (\bibinfo {year} {2012})}\BibitemShut {NoStop}%
\bibitem [{\citenamefont {M\"uller}\ \emph {et~al.}(2009)\citenamefont
  {M\"uller}, \citenamefont {Lesanovsky}, \citenamefont {Weimer}, \citenamefont
  {B\"uchler},\ and\ \citenamefont {Zoller}}]{rydberggate}%
  \BibitemOpen
  \bibfield  {author} {\bibinfo {author} {\bibfnamefont {M.}~\bibnamefont
  {M\"uller}}, \bibinfo {author} {\bibfnamefont {I.}~\bibnamefont
  {Lesanovsky}}, \bibinfo {author} {\bibfnamefont {H.}~\bibnamefont {Weimer}},
  \bibinfo {author} {\bibfnamefont {H.~P.}\ \bibnamefont {B\"uchler}}, \ and\
  \bibinfo {author} {\bibfnamefont {P.}~\bibnamefont {Zoller}},\ }\href
  {\doibase 10.1103/PhysRevLett.102.170502} {\bibfield  {journal} {\bibinfo
  {journal} {Phys. Rev. Lett.}\ }\textbf {\bibinfo {volume} {102}},\ \bibinfo
  {pages} {170502} (\bibinfo {year} {2009})}\BibitemShut {NoStop}%
\bibitem [{\citenamefont {{Jiang}}\ \emph {et~al.}(2008)\citenamefont
  {{Jiang}}, \citenamefont {{Brennen}}, \citenamefont {{Gorshkov}},
  \citenamefont {{Hammerer}}, \citenamefont {{Hafezi}}, \citenamefont
  {{Demler}}, \citenamefont {{Lukin}},\ and\ \citenamefont
  {{Zoller}}}]{cavitygate}%
  \BibitemOpen
  \bibfield  {author} {\bibinfo {author} {\bibfnamefont {L.}~\bibnamefont
  {{Jiang}}}, \bibinfo {author} {\bibfnamefont {G.~K.}\ \bibnamefont
  {{Brennen}}}, \bibinfo {author} {\bibfnamefont {A.~V.}\ \bibnamefont
  {{Gorshkov}}}, \bibinfo {author} {\bibfnamefont {K.}~\bibnamefont
  {{Hammerer}}}, \bibinfo {author} {\bibfnamefont {M.}~\bibnamefont
  {{Hafezi}}}, \bibinfo {author} {\bibfnamefont {E.}~\bibnamefont {{Demler}}},
  \bibinfo {author} {\bibfnamefont {M.~D.}\ \bibnamefont {{Lukin}}}, \ and\
  \bibinfo {author} {\bibfnamefont {P.}~\bibnamefont {{Zoller}}},\ }\href
  {\doibase 10.1038/nphys943} {\bibfield  {journal} {\bibinfo  {journal}
  {Nature Physics}\ }\textbf {\bibinfo {volume} {4}},\ \bibinfo {pages} {482}
  (\bibinfo {year} {2008})},\ \Eprint {http://arxiv.org/abs/0711.1365}
  {arXiv:0711.1365 [quant-ph]} \BibitemShut {NoStop}%
\bibitem [{\citenamefont {Davis}\ \emph {et~al.}(2016)\citenamefont {Davis},
  \citenamefont {Bentsen},\ and\ \citenamefont {Schleier-Smith}}]{Davis15}%
  \BibitemOpen
  \bibfield  {author} {\bibinfo {author} {\bibfnamefont {E.}~\bibnamefont
  {Davis}}, \bibinfo {author} {\bibfnamefont {G.}~\bibnamefont {Bentsen}}, \
  and\ \bibinfo {author} {\bibfnamefont {M.}~\bibnamefont {Schleier-Smith}},\
  }\href {\doibase 10.1103/PhysRevLett.116.053601} {\bibfield  {journal}
  {\bibinfo  {journal} {Phys. Rev. Lett.}\ }\textbf {\bibinfo {volume} {116}},\
  \bibinfo {pages} {053601} (\bibinfo {year} {2016})}\BibitemShut {NoStop}%
\bibitem [{\citenamefont {Sachdev}\ and\ \citenamefont {Ye}(1993)}]{sachdevye}%
  \BibitemOpen
  \bibfield  {author} {\bibinfo {author} {\bibfnamefont {S.}~\bibnamefont
  {Sachdev}}\ and\ \bibinfo {author} {\bibfnamefont {J.}~\bibnamefont {Ye}},\
  }\href {\doibase 10.1103/PhysRevLett.70.3339} {\bibfield  {journal} {\bibinfo
   {journal} {Phys. Rev. Lett.}\ }\textbf {\bibinfo {volume} {70}},\ \bibinfo
  {pages} {3339} (\bibinfo {year} {1993})}\BibitemShut {NoStop}%
\bibitem [{\citenamefont {Dowling}\ \emph {et~al.}(1994)\citenamefont
  {Dowling}, \citenamefont {Agarwal},\ and\ \citenamefont
  {Schleich}}]{Dowling94}%
  \BibitemOpen
  \bibfield  {author} {\bibinfo {author} {\bibfnamefont {J.~P.}\ \bibnamefont
  {Dowling}}, \bibinfo {author} {\bibfnamefont {G.~S.}\ \bibnamefont
  {Agarwal}}, \ and\ \bibinfo {author} {\bibfnamefont {W.~P.}\ \bibnamefont
  {Schleich}},\ }\href {\doibase 10.1103/PhysRevA.49.4101} {\bibfield
  {journal} {\bibinfo  {journal} {Phys. Rev. A}\ }\textbf {\bibinfo {volume}
  {49}},\ \bibinfo {pages} {4101} (\bibinfo {year} {1994})}\BibitemShut
  {NoStop}%
\bibitem [{\citenamefont {Haake}\ \emph {et~al.}(1987)\citenamefont {Haake},
  \citenamefont {Ku{\'{s}}},\ and\ \citenamefont {Scharf}}]{haake}%
  \BibitemOpen
  \bibfield  {author} {\bibinfo {author} {\bibfnamefont {F.}~\bibnamefont
  {Haake}}, \bibinfo {author} {\bibfnamefont {M.}~\bibnamefont {Ku{\'{s}}}}, \
  and\ \bibinfo {author} {\bibfnamefont {R.}~\bibnamefont {Scharf}},\ }\href
  {\doibase 10.1007/BF01303727} {\bibfield  {journal} {\bibinfo  {journal}
  {Zeitschrift f{\"u}r Physik B Condensed Matter}\ }\textbf {\bibinfo {volume}
  {65}},\ \bibinfo {pages} {381} (\bibinfo {year} {1987})}\BibitemShut
  {NoStop}%
\bibitem [{\citenamefont {{Chaudhury}}\ \emph {et~al.}(2009)\citenamefont
  {{Chaudhury}}, \citenamefont {{Smith}}, \citenamefont {{Anderson}},
  \citenamefont {{Ghose}},\ and\ \citenamefont
  {{Jessen}}}]{2009Natur.461..768C}%
  \BibitemOpen
  \bibfield  {author} {\bibinfo {author} {\bibfnamefont {S.}~\bibnamefont
  {{Chaudhury}}}, \bibinfo {author} {\bibfnamefont {A.}~\bibnamefont
  {{Smith}}}, \bibinfo {author} {\bibfnamefont {B.~E.}\ \bibnamefont
  {{Anderson}}}, \bibinfo {author} {\bibfnamefont {S.}~\bibnamefont {{Ghose}}},
  \ and\ \bibinfo {author} {\bibfnamefont {P.~S.}\ \bibnamefont {{Jessen}}},\
  }\href {\doibase 10.1038/nature08396} {\bibfield  {journal} {\bibinfo
  {journal} {\nat}\ }\textbf {\bibinfo {volume} {461}},\ \bibinfo {pages} {768}
  (\bibinfo {year} {2009})}\BibitemShut {NoStop}%
\bibitem [{\citenamefont {{Wang}}\ \emph {et~al.}(2004)\citenamefont {{Wang}},
  \citenamefont {{Ghose}}, \citenamefont {{Sanders}},\ and\ \citenamefont
  {{Hu}}}]{2004PhRvE..70a6217W}%
  \BibitemOpen
  \bibfield  {author} {\bibinfo {author} {\bibfnamefont {X.}~\bibnamefont
  {{Wang}}}, \bibinfo {author} {\bibfnamefont {S.}~\bibnamefont {{Ghose}}},
  \bibinfo {author} {\bibfnamefont {B.~C.}\ \bibnamefont {{Sanders}}}, \ and\
  \bibinfo {author} {\bibfnamefont {B.}~\bibnamefont {{Hu}}},\ }\href {\doibase
  10.1103/PhysRevE.70.016217} {\bibfield  {journal} {\bibinfo  {journal}
  {\pre}\ }\textbf {\bibinfo {volume} {70}},\ \bibinfo {eid} {016217} (\bibinfo
  {year} {2004})},\ \Eprint {http://arxiv.org/abs/quant-ph/0312047}
  {quant-ph/0312047} \BibitemShut {NoStop}%
\bibitem [{SM()}]{SM}%
  \BibitemOpen
  \href@noop {} {\ }\bibinfo {note} {See Supplemental Material at [URL will be
  inserted by publisher] for additional background and supporting
  derivations.}\BibitemShut {Stop}%
\bibitem [{\citenamefont {Colombe}\ \emph {et~al.}(2007)\citenamefont
  {Colombe}, \citenamefont {Steinmetz}, \citenamefont {Dubois}, \citenamefont
  {Linke}, \citenamefont {Hunger},\ and\ \citenamefont {Reichel}}]{Colombe07}%
  \BibitemOpen
  \bibfield  {author} {\bibinfo {author} {\bibfnamefont {Y.}~\bibnamefont
  {Colombe}}, \bibinfo {author} {\bibfnamefont {T.}~\bibnamefont {Steinmetz}},
  \bibinfo {author} {\bibfnamefont {G.}~\bibnamefont {Dubois}}, \bibinfo
  {author} {\bibfnamefont {F.}~\bibnamefont {Linke}}, \bibinfo {author}
  {\bibfnamefont {D.}~\bibnamefont {Hunger}}, \ and\ \bibinfo {author}
  {\bibfnamefont {J.}~\bibnamefont {Reichel}},\ }\href@noop {} {\bibfield
  {journal} {\bibinfo  {journal} {Nature}\ }\textbf {\bibinfo {volume} {450}},\
  \bibinfo {pages} {272} (\bibinfo {year} {2007})}\BibitemShut {NoStop}%
\bibitem [{\citenamefont {Klinder}\ \emph {et~al.}(2015)\citenamefont
  {Klinder}, \citenamefont {Ke\ss{}ler}, \citenamefont {Bakhtiari},
  \citenamefont {Thorwart},\ and\ \citenamefont {Hemmerich}}]{Klinder15}%
  \BibitemOpen
  \bibfield  {author} {\bibinfo {author} {\bibfnamefont {J.}~\bibnamefont
  {Klinder}}, \bibinfo {author} {\bibfnamefont {H.}~\bibnamefont {Ke\ss{}ler}},
  \bibinfo {author} {\bibfnamefont {M.~R.}\ \bibnamefont {Bakhtiari}}, \bibinfo
  {author} {\bibfnamefont {M.}~\bibnamefont {Thorwart}}, \ and\ \bibinfo
  {author} {\bibfnamefont {A.}~\bibnamefont {Hemmerich}},\ }\href {\doibase
  10.1103/PhysRevLett.115.230403} {\bibfield  {journal} {\bibinfo  {journal}
  {Phys. Rev. Lett.}\ }\textbf {\bibinfo {volume} {115}},\ \bibinfo {pages}
  {230403} (\bibinfo {year} {2015})}\BibitemShut {NoStop}%
\bibitem [{\citenamefont {{Kitaev}}(2015)}]{kitaev4f}%
  \BibitemOpen
  \bibfield  {author} {\bibinfo {author} {\bibfnamefont {A.}~\bibnamefont
  {{Kitaev}}},\ }\href {http://online.kitp.ucsb.edu/online/entangled15/kitaev/}
  {\bibfield  {journal} {\bibinfo  {journal} {talk at KITP Santa Barbara}\ }
  (\bibinfo {year} {2015})}\BibitemShut {NoStop}%
\bibitem [{\citenamefont {D'Alessio}\ and\ \citenamefont
  {Rigol}(2014)}]{Rigol14}%
  \BibitemOpen
  \bibfield  {author} {\bibinfo {author} {\bibfnamefont {L.}~\bibnamefont
  {D'Alessio}}\ and\ \bibinfo {author} {\bibfnamefont {M.}~\bibnamefont
  {Rigol}},\ }\href {\doibase 10.1103/PhysRevX.4.041048} {\bibfield  {journal}
  {\bibinfo  {journal} {Phys. Rev. X}\ }\textbf {\bibinfo {volume} {4}},\
  \bibinfo {pages} {041048} (\bibinfo {year} {2014})}\BibitemShut {NoStop}%
\bibitem [{\citenamefont {{Danshita}}\ \emph {et~al.}(2016)\citenamefont
  {{Danshita}}, \citenamefont {{Hanada}},\ and\ \citenamefont
  {{Tezuka}}}]{2016arXiv160602454D}%
  \BibitemOpen
  \bibfield  {author} {\bibinfo {author} {\bibfnamefont {I.}~\bibnamefont
  {{Danshita}}}, \bibinfo {author} {\bibfnamefont {M.}~\bibnamefont
  {{Hanada}}}, \ and\ \bibinfo {author} {\bibfnamefont {M.}~\bibnamefont
  {{Tezuka}}},\ }\href@noop {} {\bibfield  {journal} {\bibinfo  {journal}
  {ArXiv e-prints}\ } (\bibinfo {year} {2016})},\ \Eprint
  {http://arxiv.org/abs/1606.02454} {arXiv:1606.02454 [cond-mat.quant-gas]}
  \BibitemShut {NoStop}%
\bibitem [{\citenamefont {Anderson}(1958)}]{PhysRev.109.1492}%
  \BibitemOpen
  \bibfield  {author} {\bibinfo {author} {\bibfnamefont {P.~W.}\ \bibnamefont
  {Anderson}},\ }\href {\doibase 10.1103/PhysRev.109.1492} {\bibfield
  {journal} {\bibinfo  {journal} {Phys. Rev.}\ }\textbf {\bibinfo {volume}
  {109}},\ \bibinfo {pages} {1492} (\bibinfo {year} {1958})}\BibitemShut
  {NoStop}%
\bibitem [{\citenamefont {Billy}\ \emph {et~al.}(2008)\citenamefont {Billy},
  \citenamefont {Josse}, \citenamefont {Zuo}, \citenamefont {Bernard},
  \citenamefont {Hambrecht}, \citenamefont {Lugan}, \citenamefont
  {Cl{\'e}ment}, \citenamefont {Sanchez-Palencia}, \citenamefont {Bouyer},\
  and\ \citenamefont {Aspect}}]{billy2008direct}%
  \BibitemOpen
  \bibfield  {author} {\bibinfo {author} {\bibfnamefont {J.}~\bibnamefont
  {Billy}}, \bibinfo {author} {\bibfnamefont {V.}~\bibnamefont {Josse}},
  \bibinfo {author} {\bibfnamefont {Z.}~\bibnamefont {Zuo}}, \bibinfo {author}
  {\bibfnamefont {A.}~\bibnamefont {Bernard}}, \bibinfo {author} {\bibfnamefont
  {B.}~\bibnamefont {Hambrecht}}, \bibinfo {author} {\bibfnamefont
  {P.}~\bibnamefont {Lugan}}, \bibinfo {author} {\bibfnamefont
  {D.}~\bibnamefont {Cl{\'e}ment}}, \bibinfo {author} {\bibfnamefont
  {L.}~\bibnamefont {Sanchez-Palencia}}, \bibinfo {author} {\bibfnamefont
  {P.}~\bibnamefont {Bouyer}}, \ and\ \bibinfo {author} {\bibfnamefont
  {A.}~\bibnamefont {Aspect}},\ }\href@noop {} {\bibfield  {journal} {\bibinfo
  {journal} {Nature}\ }\textbf {\bibinfo {volume} {453}},\ \bibinfo {pages}
  {891} (\bibinfo {year} {2008})}\BibitemShut {NoStop}%
\bibitem [{\citenamefont {Roati}\ \emph {et~al.}(2008)\citenamefont {Roati},
  \citenamefont {D’Errico}, \citenamefont {Fallani}, \citenamefont {Fattori},
  \citenamefont {Fort}, \citenamefont {Zaccanti}, \citenamefont {Modugno},
  \citenamefont {Modugno},\ and\ \citenamefont {Inguscio}}]{roati2008anderson}%
  \BibitemOpen
  \bibfield  {author} {\bibinfo {author} {\bibfnamefont {G.}~\bibnamefont
  {Roati}}, \bibinfo {author} {\bibfnamefont {C.}~\bibnamefont {D’Errico}},
  \bibinfo {author} {\bibfnamefont {L.}~\bibnamefont {Fallani}}, \bibinfo
  {author} {\bibfnamefont {M.}~\bibnamefont {Fattori}}, \bibinfo {author}
  {\bibfnamefont {C.}~\bibnamefont {Fort}}, \bibinfo {author} {\bibfnamefont
  {M.}~\bibnamefont {Zaccanti}}, \bibinfo {author} {\bibfnamefont
  {G.}~\bibnamefont {Modugno}}, \bibinfo {author} {\bibfnamefont
  {M.}~\bibnamefont {Modugno}}, \ and\ \bibinfo {author} {\bibfnamefont
  {M.}~\bibnamefont {Inguscio}},\ }\href@noop {} {\bibfield  {journal}
  {\bibinfo  {journal} {Nature}\ }\textbf {\bibinfo {volume} {453}},\ \bibinfo
  {pages} {895} (\bibinfo {year} {2008})}\BibitemShut {NoStop}%
\bibitem [{\citenamefont {{Basko}}\ \emph {et~al.}(2006)\citenamefont
  {{Basko}}, \citenamefont {{Aleiner}},\ and\ \citenamefont
  {{Altshuler}}}]{2006AnPhy.321.1126B}%
  \BibitemOpen
  \bibfield  {author} {\bibinfo {author} {\bibfnamefont {D.~M.}\ \bibnamefont
  {{Basko}}}, \bibinfo {author} {\bibfnamefont {I.~L.}\ \bibnamefont
  {{Aleiner}}}, \ and\ \bibinfo {author} {\bibfnamefont {B.~L.}\ \bibnamefont
  {{Altshuler}}},\ }\href {\doibase 10.1016/j.aop.2005.11.014} {\bibfield
  {journal} {\bibinfo  {journal} {Annals of Physics}\ }\textbf {\bibinfo
  {volume} {321}},\ \bibinfo {pages} {1126} (\bibinfo {year} {2006})},\ \Eprint
  {http://arxiv.org/abs/cond-mat/0506617} {cond-mat/0506617} \BibitemShut
  {NoStop}%
\bibitem [{\citenamefont {Schreiber}\ \emph {et~al.}(2015)\citenamefont
  {Schreiber}, \citenamefont {Hodgman}, \citenamefont {Bordia}, \citenamefont
  {L{\"u}schen}, \citenamefont {Fischer}, \citenamefont {Vosk}, \citenamefont
  {Altman}, \citenamefont {Schneider},\ and\ \citenamefont
  {Bloch}}]{Schreiber842}%
  \BibitemOpen
  \bibfield  {author} {\bibinfo {author} {\bibfnamefont {M.}~\bibnamefont
  {Schreiber}}, \bibinfo {author} {\bibfnamefont {S.~S.}\ \bibnamefont
  {Hodgman}}, \bibinfo {author} {\bibfnamefont {P.}~\bibnamefont {Bordia}},
  \bibinfo {author} {\bibfnamefont {H.~P.}\ \bibnamefont {L{\"u}schen}},
  \bibinfo {author} {\bibfnamefont {M.~H.}\ \bibnamefont {Fischer}}, \bibinfo
  {author} {\bibfnamefont {R.}~\bibnamefont {Vosk}}, \bibinfo {author}
  {\bibfnamefont {E.}~\bibnamefont {Altman}}, \bibinfo {author} {\bibfnamefont
  {U.}~\bibnamefont {Schneider}}, \ and\ \bibinfo {author} {\bibfnamefont
  {I.}~\bibnamefont {Bloch}},\ }\href@noop {} {\bibfield  {journal} {\bibinfo
  {journal} {Science}\ }\textbf {\bibinfo {volume} {349}},\ \bibinfo {pages}
  {842} (\bibinfo {year} {2015})}\BibitemShut {NoStop}%
\bibitem [{\citenamefont {Bohnet}\ \emph {et~al.}(2015)\citenamefont {Bohnet},
  \citenamefont {Sawyer}, \citenamefont {Britton}, \citenamefont {Wall},
  \citenamefont {Rey}, \citenamefont {Foss-Feig},\ and\ \citenamefont
  {Bollinger}}]{bohnet2015quantum}%
  \BibitemOpen
  \bibfield  {author} {\bibinfo {author} {\bibfnamefont {J.~G.}\ \bibnamefont
  {Bohnet}}, \bibinfo {author} {\bibfnamefont {B.~C.}\ \bibnamefont {Sawyer}},
  \bibinfo {author} {\bibfnamefont {J.~W.}\ \bibnamefont {Britton}}, \bibinfo
  {author} {\bibfnamefont {M.~L.}\ \bibnamefont {Wall}}, \bibinfo {author}
  {\bibfnamefont {A.~M.}\ \bibnamefont {Rey}}, \bibinfo {author} {\bibfnamefont
  {M.}~\bibnamefont {Foss-Feig}}, \ and\ \bibinfo {author} {\bibfnamefont
  {J.~J.}\ \bibnamefont {Bollinger}},\ }\href@noop {} {\bibfield  {journal}
  {\bibinfo  {journal} {arXiv preprint arXiv:1512.03756}\ } (\bibinfo {year}
  {2015})}\BibitemShut {NoStop}%
\bibitem [{\citenamefont {van Bijnen}\ and\ \citenamefont
  {Pohl}(2015)}]{vanBijnen15}%
  \BibitemOpen
  \bibfield  {author} {\bibinfo {author} {\bibfnamefont {R.}~\bibnamefont {van
  Bijnen}}\ and\ \bibinfo {author} {\bibfnamefont {T.}~\bibnamefont {Pohl}},\
  }\href@noop {} {\bibfield  {journal} {\bibinfo  {journal} {Phys. Rev. Lett.}\
  }\textbf {\bibinfo {volume} {114}},\ \bibinfo {pages} {243002} (\bibinfo
  {year} {2015})}\BibitemShut {NoStop}%
\bibitem [{\citenamefont {Glaetzle}\ \emph {et~al.}(2015)\citenamefont
  {Glaetzle}, \citenamefont {Dalmonte}, \citenamefont {Nath}, \citenamefont
  {Gross}, \citenamefont {Bloch},\ and\ \citenamefont {Zoller}}]{Glaetzle15}%
  \BibitemOpen
  \bibfield  {author} {\bibinfo {author} {\bibfnamefont {A.~W.}\ \bibnamefont
  {Glaetzle}}, \bibinfo {author} {\bibfnamefont {M.}~\bibnamefont {Dalmonte}},
  \bibinfo {author} {\bibfnamefont {R.}~\bibnamefont {Nath}}, \bibinfo {author}
  {\bibfnamefont {C.}~\bibnamefont {Gross}}, \bibinfo {author} {\bibfnamefont
  {I.}~\bibnamefont {Bloch}}, \ and\ \bibinfo {author} {\bibfnamefont
  {P.}~\bibnamefont {Zoller}},\ }\href@noop {} {\bibfield  {journal} {\bibinfo
  {journal} {Phys. Rev. Lett.}\ }\textbf {\bibinfo {volume} {114}},\ \bibinfo
  {pages} {173002} (\bibinfo {year} {2015})}\BibitemShut {NoStop}%
\bibitem [{\citenamefont {Zeiher}\ \emph {et~al.}(2016)\citenamefont {Zeiher},
  \citenamefont {van Bijnen}, \citenamefont {Schau{\ss}}, \citenamefont {Hild},
  \citenamefont {Choi}, \citenamefont {Pohl}, \citenamefont {Bloch},\ and\
  \citenamefont {Gross}}]{Zeiher16}%
  \BibitemOpen
  \bibfield  {author} {\bibinfo {author} {\bibfnamefont {J.}~\bibnamefont
  {Zeiher}}, \bibinfo {author} {\bibfnamefont {R.}~\bibnamefont {van Bijnen}},
  \bibinfo {author} {\bibfnamefont {P.}~\bibnamefont {Schau{\ss}}}, \bibinfo
  {author} {\bibfnamefont {S.}~\bibnamefont {Hild}}, \bibinfo {author}
  {\bibfnamefont {J.-y.}\ \bibnamefont {Choi}}, \bibinfo {author}
  {\bibfnamefont {T.}~\bibnamefont {Pohl}}, \bibinfo {author} {\bibfnamefont
  {I.}~\bibnamefont {Bloch}}, \ and\ \bibinfo {author} {\bibfnamefont
  {C.}~\bibnamefont {Gross}},\ }\href@noop {} {\bibfield  {journal} {\bibinfo
  {journal} {arXiv:1602.06313}\ } (\bibinfo {year} {2016})}\BibitemShut
  {NoStop}%
\bibitem [{\citenamefont {Struck}\ \emph {et~al.}(2012)\citenamefont {Struck},
  \citenamefont {\"Olschl\"ager}, \citenamefont {Weinberg}, \citenamefont
  {Hauke}, \citenamefont {Simonet}, \citenamefont {Eckardt}, \citenamefont
  {Lewenstein}, \citenamefont {Sengstock},\ and\ \citenamefont
  {Windpassinger}}]{shakegeneralproposal}%
  \BibitemOpen
  \bibfield  {author} {\bibinfo {author} {\bibfnamefont {J.}~\bibnamefont
  {Struck}}, \bibinfo {author} {\bibfnamefont {C.}~\bibnamefont
  {\"Olschl\"ager}}, \bibinfo {author} {\bibfnamefont {M.}~\bibnamefont
  {Weinberg}}, \bibinfo {author} {\bibfnamefont {P.}~\bibnamefont {Hauke}},
  \bibinfo {author} {\bibfnamefont {J.}~\bibnamefont {Simonet}}, \bibinfo
  {author} {\bibfnamefont {A.}~\bibnamefont {Eckardt}}, \bibinfo {author}
  {\bibfnamefont {M.}~\bibnamefont {Lewenstein}}, \bibinfo {author}
  {\bibfnamefont {K.}~\bibnamefont {Sengstock}}, \ and\ \bibinfo {author}
  {\bibfnamefont {P.}~\bibnamefont {Windpassinger}},\ }\href {\doibase
  10.1103/PhysRevLett.108.225304} {\bibfield  {journal} {\bibinfo  {journal}
  {Phys. Rev. Lett.}\ }\textbf {\bibinfo {volume} {108}},\ \bibinfo {pages}
  {225304} (\bibinfo {year} {2012})}\BibitemShut {NoStop}%
\bibitem [{\citenamefont {Struck}\ \emph {et~al.}(2011)\citenamefont {Struck},
  \citenamefont {Ölschläger}, \citenamefont {Le~Targat}, \citenamefont
  {Soltan-Panahi}, \citenamefont {Eckardt}, \citenamefont {Lewenstein},
  \citenamefont {Windpassinger},\ and\ \citenamefont {Sengstock}}]{shake4}%
  \BibitemOpen
  \bibfield  {author} {\bibinfo {author} {\bibfnamefont {J.}~\bibnamefont
  {Struck}}, \bibinfo {author} {\bibfnamefont {C.}~\bibnamefont
  {Ölschläger}}, \bibinfo {author} {\bibfnamefont {R.}~\bibnamefont
  {Le~Targat}}, \bibinfo {author} {\bibfnamefont {P.}~\bibnamefont
  {Soltan-Panahi}}, \bibinfo {author} {\bibfnamefont {A.}~\bibnamefont
  {Eckardt}}, \bibinfo {author} {\bibfnamefont {M.}~\bibnamefont {Lewenstein}},
  \bibinfo {author} {\bibfnamefont {P.}~\bibnamefont {Windpassinger}}, \ and\
  \bibinfo {author} {\bibfnamefont {K.}~\bibnamefont {Sengstock}},\ }\href
  {\doibase 10.1126/science.1207239} {\bibfield  {journal} {\bibinfo  {journal}
  {Science}\ }\textbf {\bibinfo {volume} {333}},\ \bibinfo {pages} {996}
  (\bibinfo {year} {2011})},\ \Eprint
  {http://arxiv.org/abs/http://www.sciencemag.org/content/333/6045/996.full.pdf}
  {http://www.sciencemag.org/content/333/6045/996.full.pdf} \BibitemShut
  {NoStop}%
\bibitem [{\citenamefont {Aidelsburger}\ \emph {et~al.}(2011)\citenamefont
  {Aidelsburger}, \citenamefont {Atala}, \citenamefont {Nascimb\`ene},
  \citenamefont {Trotzky}, \citenamefont {Chen},\ and\ \citenamefont
  {Bloch}}]{raman_bloch}%
  \BibitemOpen
  \bibfield  {author} {\bibinfo {author} {\bibfnamefont {M.}~\bibnamefont
  {Aidelsburger}}, \bibinfo {author} {\bibfnamefont {M.}~\bibnamefont {Atala}},
  \bibinfo {author} {\bibfnamefont {S.}~\bibnamefont {Nascimb\`ene}}, \bibinfo
  {author} {\bibfnamefont {S.}~\bibnamefont {Trotzky}}, \bibinfo {author}
  {\bibfnamefont {Y.-A.}\ \bibnamefont {Chen}}, \ and\ \bibinfo {author}
  {\bibfnamefont {I.}~\bibnamefont {Bloch}},\ }\href {\doibase
  10.1103/PhysRevLett.107.255301} {\bibfield  {journal} {\bibinfo  {journal}
  {Phys. Rev. Lett.}\ }\textbf {\bibinfo {volume} {107}},\ \bibinfo {pages}
  {255301} (\bibinfo {year} {2011})}\BibitemShut {NoStop}%
\bibitem [{\citenamefont {Bakr}\ \emph {et~al.}(2009)\citenamefont {Bakr},
  \citenamefont {Gillen}, \citenamefont {Peng}, \citenamefont {F{\"o}lling},\
  and\ \citenamefont {Greiner}}]{bakr2009quantum}%
  \BibitemOpen
  \bibfield  {author} {\bibinfo {author} {\bibfnamefont {W.~S.}\ \bibnamefont
  {Bakr}}, \bibinfo {author} {\bibfnamefont {J.~I.}\ \bibnamefont {Gillen}},
  \bibinfo {author} {\bibfnamefont {A.}~\bibnamefont {Peng}}, \bibinfo {author}
  {\bibfnamefont {S.}~\bibnamefont {F{\"o}lling}}, \ and\ \bibinfo {author}
  {\bibfnamefont {M.}~\bibnamefont {Greiner}},\ }\href@noop {} {\bibfield
  {journal} {\bibinfo  {journal} {Nature}\ }\textbf {\bibinfo {volume} {462}},\
  \bibinfo {pages} {74} (\bibinfo {year} {2009})}\BibitemShut {NoStop}%
\bibitem [{\citenamefont {Weitenberg}\ \emph {et~al.}(2011)\citenamefont
  {Weitenberg}, \citenamefont {Endres}, \citenamefont {Sherson}, \citenamefont
  {Cheneau}, \citenamefont {Schauss}, \citenamefont {Fukuhara}, \citenamefont
  {Bloch},\ and\ \citenamefont {Kuhr}}]{weitenberg2011single}%
  \BibitemOpen
  \bibfield  {author} {\bibinfo {author} {\bibfnamefont {C.}~\bibnamefont
  {Weitenberg}}, \bibinfo {author} {\bibfnamefont {M.}~\bibnamefont {Endres}},
  \bibinfo {author} {\bibfnamefont {J.~F.}\ \bibnamefont {Sherson}}, \bibinfo
  {author} {\bibfnamefont {M.}~\bibnamefont {Cheneau}}, \bibinfo {author}
  {\bibfnamefont {P.}~\bibnamefont {Schauss}}, \bibinfo {author} {\bibfnamefont
  {T.}~\bibnamefont {Fukuhara}}, \bibinfo {author} {\bibfnamefont
  {I.}~\bibnamefont {Bloch}}, \ and\ \bibinfo {author} {\bibfnamefont
  {S.}~\bibnamefont {Kuhr}},\ }\href@noop {} {\bibfield  {journal} {\bibinfo
  {journal} {Nature}\ }\textbf {\bibinfo {volume} {471}},\ \bibinfo {pages}
  {319} (\bibinfo {year} {2011})}\BibitemShut {NoStop}%
\bibitem [{\citenamefont {Daley}(2014)}]{Daley:2014fha}%
  \BibitemOpen
  \bibfield  {author} {\bibinfo {author} {\bibfnamefont {A.~J.}\ \bibnamefont
  {Daley}},\ }\href {\doibase 10.1080/00018732.2014.933502} {\bibfield
  {journal} {\bibinfo  {journal} {Adv. Phys.}\ }\textbf {\bibinfo {volume}
  {63}},\ \bibinfo {pages} {77} (\bibinfo {year} {2014})},\ \Eprint
  {http://arxiv.org/abs/1405.6694} {arXiv:1405.6694 [quant-ph]} \BibitemShut
  {NoStop}%
\bibitem [{\citenamefont {Reiter}\ and\ \citenamefont
  {S\o{}rensen}(2012)}]{Reiter}%
  \BibitemOpen
  \bibfield  {author} {\bibinfo {author} {\bibfnamefont {F.}~\bibnamefont
  {Reiter}}\ and\ \bibinfo {author} {\bibfnamefont {A.~S.}\ \bibnamefont
  {S\o{}rensen}},\ }\href {\doibase 10.1103/PhysRevA.85.032111} {\bibfield
  {journal} {\bibinfo  {journal} {Phys. Rev. A}\ }\textbf {\bibinfo {volume}
  {85}},\ \bibinfo {pages} {032111} (\bibinfo {year} {2012})}\BibitemShut
  {NoStop}%
\bibitem [{\citenamefont {Jiang}\ \emph {et~al.}(2008)\citenamefont {Jiang},
  \citenamefont {Brennen}, \citenamefont {Gorshkov}, \citenamefont {Hammerer},
  \citenamefont {Hafezi}, \citenamefont {Demler}, \citenamefont {Lukin},\ and\
  \citenamefont {Zoller}}]{Jiang}%
  \BibitemOpen
  \bibfield  {author} {\bibinfo {author} {\bibfnamefont {L.}~\bibnamefont
  {Jiang}}, \bibinfo {author} {\bibfnamefont {G.~K.}\ \bibnamefont {Brennen}},
  \bibinfo {author} {\bibfnamefont {A.~V.}\ \bibnamefont {Gorshkov}}, \bibinfo
  {author} {\bibfnamefont {K.}~\bibnamefont {Hammerer}}, \bibinfo {author}
  {\bibfnamefont {M.}~\bibnamefont {Hafezi}}, \bibinfo {author} {\bibfnamefont
  {E.}~\bibnamefont {Demler}}, \bibinfo {author} {\bibfnamefont {M.~D.}\
  \bibnamefont {Lukin}}, \ and\ \bibinfo {author} {\bibfnamefont
  {P.}~\bibnamefont {Zoller}},\ }\href {\doibase 10.1038/nphys943} {\bibfield
  {journal} {\bibinfo  {journal} {Nat Phys}\ }\textbf {\bibinfo {volume} {4}}
  (\bibinfo {year} {2008}),\ 10.1038/nphys943}\BibitemShut {NoStop}%
\bibitem [{\citenamefont {Schleier-Smith}\ \emph {et~al.}(2010)\citenamefont
  {Schleier-Smith}, \citenamefont {Leroux},\ and\ \citenamefont
  {Vuleti{\'c}}}]{SchleierSmith10a}%
  \BibitemOpen
  \bibfield  {author} {\bibinfo {author} {\bibfnamefont {M.~H.}\ \bibnamefont
  {Schleier-Smith}}, \bibinfo {author} {\bibfnamefont {I.~D.}\ \bibnamefont
  {Leroux}}, \ and\ \bibinfo {author} {\bibfnamefont {V.}~\bibnamefont
  {Vuleti{\'c}}},\ }\href {\doibase 10.1103/PhysRevA.81.021804} {\bibfield
  {journal} {\bibinfo  {journal} {Phys. Rev. A}\ }\textbf {\bibinfo {volume}
  {81}},\ \bibinfo {pages} {021804} (\bibinfo {year} {2010})}\BibitemShut
  {NoStop}%
\end{thebibliography}%

\section{Measuring time-ordered correlation functions}
\label{sec:timeorder}

Here we demonstrate that time-ordered correlation functions can be measured using only forward time evolution.

Suppose we want to measure a correlation function of the form
\beq
G(t,s) = \left\langle\left(V_n(t_n) ... V_1(t_1)\right)^\dagger \left(W_n(s_n) ... W_1(s_1)\right) \right\rangle
\eeq
where $V_i$ and $W_j$ are unitary operators (which may or may not all be distinct) acting on some system $A$ and  $\{t_i\}$ and $\{s_i\}$ are nondecreasing time sequences. First, observe that $G$ is a time ordered correlation function. All the $W$ and $V$ operators are manifestly in time order within their respective parenthesis and the $\dagger$ on the $V$ operators reverses the order of all such operators. Hence the time labels of the operators in $G$ first increase ($s$ part) and then decrease ($t$ part).

Note that by choosing some of the $V_i$ or $W_j$ to be identity operators we may assume without loss of generality that $\{s_i\}$ and $\{t_j\}$ have the same number of points and that $t_i = s_i$. Then we proceed as in the general protocol above by introducing a control qubit $\CC$ and initializing the total system into the state
\beq
\ket{\text{initial}} = |\psi\rangle_\CS \frac{\ket{0}_\CC + \ket{1}_\CC}{\sqrt{2}}.
\eeq
Then apply the gate sequence
\begin{itemize}
\item[1.] $e^{-i t_1 H}_\CS \otimes I_\CC$
\item[2.] $(W_1)_\CS \otimes \ket{0}\bra{0}_\CC + (V_1)_\CS \otimes \ket{1}\bra{1}_\CC$
\item[3.] $e^{-i (t_2 - t_1) H}_\CS \otimes I_\CC$
\item[4.] $(W_2)_\CS \otimes \ket{0}\bra{0}_\CC + (V_2)_\CS \otimes \ket{1}\bra{1}_\CC$
\item[j.] ...
\item[2n-1.] $e^{-i (t_{n} -t_{n-1}) H}_\CS \otimes I_\CC$
\item[2n.] $(W_n)_\CS \otimes \ket{0}\bra{0}_\CC + (V_n)_\CS \otimes \ket{1}\bra{1}_\CC$
\end{itemize}
to produce the final state
\bea
&& |\text{final}\rangle = \nonumber \\
&&  \frac{( e^{-i t_n H} W_n(t_n) ... W_1(t_1) \ket{\psi})_\CS \ket{0}_\CC }{\sqrt{2}} \nonumber \\
&& + \frac{(e^{-i t_n H} V_n(t_n) ... V_1(t_1)\ket{\psi})_\CS \ket{1}_\CC}{\sqrt{2}}.
\eea

Some comments are in order. The time evolution prefactor common to both interference paths occurs because in the above gate protocol we did not actually apply $W_n(t_n) = e^{i t_n H} W_n e^{-i t_n H}$ but rather just $W_n e^{-i t_n H}$, so this operator prefactor cancels the reverse time evolution in the definition of $W_n(t_n)$ and $V_n(t_n)$. However, because it is common to both interference paths it has no effect on the interferometer and we may as well work with $W_n(t_n)$ and $V_n(t_n)$ as indicated. Furthermore, despite the formal appearance of factors of $U(-t_i)$ in the definition of the various $V_i(t_i)$, etc., no backwards time evolution is ever performed. The forward time evolution involved in defining $V_{k+1}(t_{k+1})$ always cancels the backwards time evolution in $V_k(t_k)$ so that only net forward evolution is required.

The final result is
\beq
\langle \text{final} | X_\CC |\text{final}\rangle = \Re(G).
\eeq
This justifies the claims that time ordered correlation functions can always be measured without reversing time. Of course, this is not to imply that the above interferometric protocol is the best way to measure $G$; all we claim is that the measurement is possible without reversing time.

\section{Out-of-time-order correlators of Hermitian operators}

If we are interested in Hermitian operators instead of unitary operators, there is a more complicated general protocol. The idea is to consider unitaries perturbed from the identity by the hermitian operator of interest (times a small coefficient). By coupling in additional marker or flag qubits which are flipped when the hermitian operator acts on the system and post selecting on measurements where the marker qubit has been flipped we can access to states where the hermitian operator has been applied. In brief, if $V_{A\, \text{flag}} = e^{i \epsilon O_A \otimes X_{\text{flag}}}$ then $V_{\CS\,\text{flag}} \ket{\psi}_\CS\ket{0}_{\text{flag}} \approx \ket{\psi}_\CS\ket{0}_{\text{flag}} + i \epsilon O_\CS \ket{\psi}_\CS \ket{1}_{\text{flag}}$ and post selecting on the flag $=1$ produces a state proportional to $O_\CS \ket{\psi}_\CS$. Of course, this is again not necessarily a very efficient way of making the measurement.

As an example, take the interferometric protocol above and introduce three additional flag qubits. The unitaries $W$ and $V$ are taken to be of the form $e^{i \epsilon O^{W,V}_\CS \otimes X_{\text{flag}}}$ and the three flag qubits are used to check that (1) the first control $V$ applies $O^V$, (2) the middle $W$ applies $O^W$, and (3) the final control $V$ applies $O^V$. Conditioned on all three flag qubits being equal to one, something which occurs with probability $\epsilon^3 (\langle O^W_t O^V O^V O^W_t \rangle + \langle O^V O_t^W O_t^W O^V \rangle )/2$, we obtain the normalized state
\beq
\frac{O^V O^W_t \ket{\psi}_A \ket{0}_\CC + O^W_t O^V\ket{\psi} \ket{1}_\CC}{\sqrt{\langle O^W_t O^V O^V O^W_t \rangle + \langle O^V O_t^W O_t^W O^V \rangle}}.
\eeq
Measurement of $X_\CC$ in this state now yields
\beq
\langle X_\CC \rangle = \frac{\langle O^V O^W_t O^V O^W_t \rangle + \langle O_t^W O^V  O_t^W O^V \rangle}{\langle O^W_t O^V O^V O^W_t \rangle + \langle O^V O_t^W O_t^W O^V \rangle};
\eeq
this is the normalized version of $\Re(F)$ for hermitian operators. Note that the normalization consists only of time ordered correlation functions.

\section{Out-of-time-order correlators in the semi-classical limit}

Here we review the basic connection between out-of-time-order correlations and chaos first observed in \cite{quasiclassicalsc}. We briefly explain the relationship between out-of-time-order correlators and the classical butterfly effect, and discuss how this intuition applies to the collective spin model studied in the main text.

Consider a single particle with position $q$ and momentum $p$ and let $q_t(q_0,p_0)$ and $p_t(q_0,p_0)$ denote the trajectories as a function of time $t$ with initial conditions $(q_0,p_0)$. If the motion is chaotic, then we expect sensitive dependence on initial conditions,
\beq
\frac{\partial q_t}{\partial q_0} \sim e^{\lambda t}
\eeq
with $\lambda$ a Lyapunov exponent. The derivative of $q_t$ with respect to the initial condition is most naturally written as a Poisson bracket,
\beq
\frac{\partial q_t}{\partial q_0} = \{ q_t,p_0\}_P,
\eeq
where the Poisson bracket is defined as
\beq
\{A,B\}_P = \frac{\partial A}{\partial q_0} \frac{\partial B}{\partial p_0} - \frac{\partial A}{\partial p_0} \frac{\partial B}{\partial q_0}.
\eeq

Quantum mechanically the Poisson bracket becomes the commutator of $q_t$ and $p_0$, and the correspondence principle gives
\beq
[q_t,p_0] \sim i\hbar \{q_t,p_0\}_P \sim i \hbar e^{\lambda t}.
\eeq
This quantity can be accessed elegantly using unitary operators (we set $q_0,p_0 \rightarrow q,p$ for notational simplicity). Let $W=e^{i a q}$ and $V=e^{i b p}$ and consider the multiplicative commutator $W_t^\dagger V^\dagger W_t V$. In the limit of small $a$ and $b$ and short time $t$ the multiplicative commutator reduces to the exponential of the usual commutator,
\beq
W_t^\dagger V^\dagger W_t V \approx e^{ i^2 a b [q_t,p]} \approx e^{ - i ab \hbar e^{\lambda t}}.
\eeq
The parameters entering this equation will depend on the state of the system as well, so a more general statement is
\beq
\langle W_t^\dagger V^\dagger W_t V  \rangle \sim e^{ - i ab \hbar e^{\lambda t}}.
\eeq
Hence the phase of this correlation function initially diverges rapidly with $t$ and, as higher order terms become important, the magnitude will also begin to decay. To the best of our knowledge, such out-of-time-order correlators were first discussed in the context of superconductivity in \cite{quasiclassicalsc}.

What is the physical meaning of this correlation function? It directly measures the overlap between two states,
\beq
e^{i H t} W e^{-i H t} V \ket{\psi_0}
\eeq
and
\beq
V e^{i H t} W e^{-i H t} \ket{\psi_0},
\eeq
and hence physically represents the sensitivity of the system to applying $V$ then $W_t$ versus applying $W_t$ then $V$. Thus it measures sensitivity to initial conditions or the ``butterfly effect". Furthermore, the key role played by reversing time, that is evolving with $e^{i H t}$ as well as with $e^{-i H t}$, should be apparent.

Supposing that the dimensionless quantity $ab \hbar = \epsilon$ is small ($1/ab$ represents a natural classical action scale in the problem), then the time required for $F$ to develop a significant phase is $\lambda^{-1} \log \(\frac{1}{\epsilon}\)$, known as the Ehrenfest time. It is the time required for the wavepacket to spread to a size of order the typical classical action. In a strongly interacting quantum system at finite temperature we expect $\lambda^{-1} \sim \tau = \frac{\hbar}{k_B T}$; in fact, this is a bound under the assumptions of \cite{chaosbound}. Thus time-ordered correlation functions, which typically decay after a time of order $\lambda^{-1}$, can decay parametrically faster than out-of-time-order correlation functions since the scale $\lambda^{-1} \log\(\frac{1}{\epsilon}\)$ is much greater than $\lambda^{-1}$ if $\epsilon$ is small.

In the kicked top model, the role of $\hbar$ is played by the inverse spin $1/S$. In the large $S$ limit the rescaled operator $\mathbf{S}/S$ becomes classical with the unit sphere interpreted as a classical phase space; $1/S$ sets the natural unit of phase space volume so that the total number of states is of order $S$. The classical dynamics on the unit sphere is chaotic and there are two Lyapunov exponents, $\pm \lambda$, which depend on the location in phase space. The analog of the Ehrenfest time is thus $\lambda^{-1} \log S$. Choosing $V$ and $W$ to be $S_z$ rotations by angle $\phi$, the scaling $\phi \sim 1/\sqrt{S}$ is designed so that at large $S$ time-ordered correlations decay on the timescale $\lambda^{-1}$ while out-of-time-order correlations decay on the longer timescale $\lambda^{-1} \log S$.

\section{Analysis of Dissipation}

\begin{figure}
\includegraphics[width=.45\textwidth]{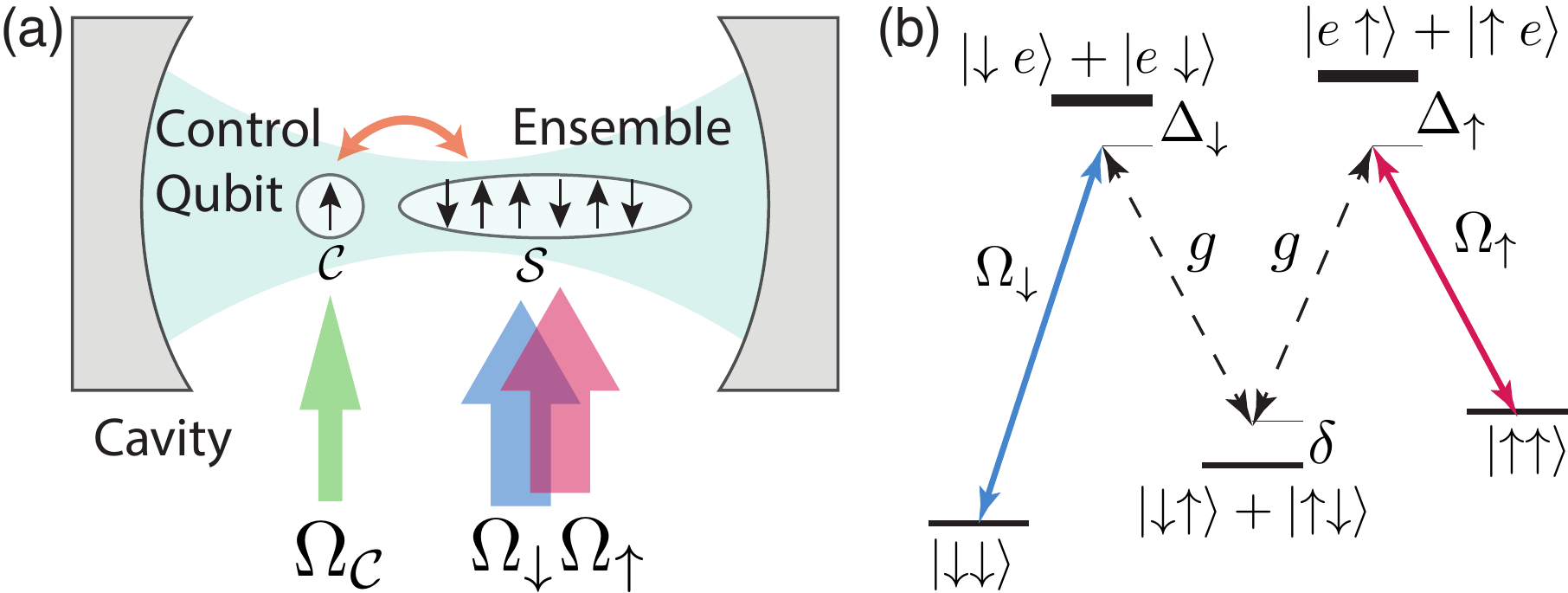}
\caption{Scheme for measuring out-of-time-order correlators. (a) Atomic ensemble $\CS$ and control qubit $\CC$ in an optical cavity are driven by control fields $\Omega_{\uparrow}, \Omega_{\downarrow}, \Omega_{\CC}$. (b) Control fields $\Omega_{\uparrow,\downarrow}$ and cavity coupling $g$ mediate pairwise interactions in the ensemble $\CS$ via 4-photon Raman transitions.}
\label{fig:cavity_setup}
\end{figure}
	
	We have used quantum trajectories methods to simulate the effects of dissipation in the cavity QED implementation. A pedagogical introduction to quantum trajectories can be found in Ref. \cite{Daley:2014fha}.\par
	As discussed in the main text, the cavity implementation suffers from cavity decay at a rate $\kappa$ and atomic spontaneous emission at a rate $\Gamma$. After adiabatically eliminating the cavity mode and excited atomic states $\ket{e}_i$ from the dynamics, these decay processes are described by a set of effective Lindblad jump operators $L$ acting on the atomic pseudo-spin states $\ket{\uparrow}_i, \ket{\downarrow}_i$ \cite{Davis15,Reiter}. The effect of cavity decay on the collective spin state is described by a Lindblad operator of the form
	\begin{equation}
		L_{\kappa} = \sqrt{\gamma} \ S_x,
	\end{equation}
	where $\gamma$ is the rate at which photons are lost per atom through the cavity mirrors. The effect of $L_{\kappa}$ is to diffusively smear out the Wigner function in directions perpendicular to $\hat{x}$. In terms of physical parameters in the cavity QED setup (Fig. \ref{fig:cavity_setup}b),
	\begin{equation}
		\label{eq:mu}
		\gamma = \kappa \left( \frac{\Omega g}{\Delta \delta} \right)^2 = \frac{2 \chi}{d},
	\end{equation}
	where the atom-atom coupling is $\chi = \Omega^2 g^2 / \Delta^2 \delta$ and we define the detuning parameter $d = 2 \delta / \kappa$. Here we have assumed uniform coupling $g_{\alpha}(r_i) = g$ and a fixed ratio between Rabi frequencies and detunings $\Omega_{\downarrow}(r_i) / \Delta_{\downarrow} = \Omega_{\uparrow}(r_i) / \Delta_{\uparrow} \equiv \Omega / \Delta$. The origin of the rate $\gamma$ can be understood in second-order perturbation theory: due to the two-photon transition driven by $\Omega_{\downarrow}, g$ (respectively $\Omega_{\uparrow}, g$), the cavity will be populated by a single photon with probability $\left( \Omega g / \Delta \delta \right)^2$ per atom, which will then leak from the cavity at a rate $\kappa$.\par
	Spontaneous emission events are described by a set of $4N$ jump operators:
	\begin{eqnarray}
		L_i^+ &=& \sqrt{\mu} \ \ket{\uparrow} \bra{\downarrow}_i \nonumber \\
		L_i^- &=& \sqrt{\mu} \ \ket{\downarrow} \bra{\uparrow}_i \nonumber \\
		L_i^{\uparrow} &=& \sqrt{\mu} \ \ket{\uparrow} \bra{\uparrow}_i \nonumber \\
		L_i^{\downarrow} &=& \sqrt{\mu} \ \ket{\downarrow} \bra{\downarrow}_i,
	\end{eqnarray}
	where $2 \mu$ is the spontaneous scattering rate per atom. These jump operators describe individual spin-flips ($L_i^+$ and $L_i^-$) and spin-projections ($L_i^{\uparrow}$ and $L_i^{\downarrow}$) induced by spontaneous scattering events. In terms of physical cavity parameters, for large two-photon detuning $d \gg 1$, we have
	\begin{equation}
		\label{eq:gamma}
		\mu \approx \frac{\Gamma}{2} \left(\frac{\Omega}{2 \Delta} \right)^2 = \frac{\chi d}{4 \eta}.
	\end{equation}
	This rate can also be understood in perturbation theory: the drive field $\Omega_{\downarrow}$ (respectively $\Omega_{\uparrow}$) will populate the excited state $\ket{e}_i$ with probability $\left( \Omega / 2 \Delta \right)^2$, and $\ket{e}_i$ will subsequently decay to the two ground states $\ket{\uparrow}_i, \ket{\downarrow}_i$ at a total rate $\Gamma$. From Eqs. \ref{eq:mu} and \ref{eq:gamma} it is evident that, once the cooperativity $\eta$ has been fixed, the detuning $d$ completely controls the relative strength of the two forms of dissipation.\par
	The cavity jump operator $L_{\kappa}$ acts symmetrically on the atomic ensemble and can be simulated numerically in the $(N+1)$-dimensional Dicke subspace of the full $2^N$-dimensional Hilbert space $\CH$. The spontaneous jump operators, however, break this symmetry and appear to require simulation of the full Hilbert space. This is prohibitive for more than just a few atoms.\par
	Fortunately, the method of quantum trajectories allows us to circumvent this problem. A single quantum trajectory consists of a known sequence of jump operators, allowing us to begin the simulation in a manageable subspace of $\CH$ and introduce new subspaces only when they are needed. The trick is to take advantage of the following identity for the addition of angular momentum:
	\begin{equation}
		\bigotimes_{i=1}^N \left(\mathbf{\frac{1}{2}}\right)_i = \bigoplus_{S}^{N/2}\bigoplus_{j=1}^{G_S} \mathbf{S}_j
	\end{equation}
	where $\mathbf{S}_j$ denotes the $(2S+1)$-dimensional Hilbert space of a spin-$S$ particle and the index $j$ labels the $G_S = \binom{N}{N/2-S}-\binom{N}{N/2-S-1}$ distinct subspaces of a given spin $S$. The parameter $S$ runs over the integers for even $N$, and over the half-integers for odd $N$. For instance, for $N = 2$ this identity gives the familiar decomposition of a pair of spin-1/2 particles into triplet and singlet subspaces: $\mathbf{\frac{1}{2}} \otimes \mathbf{\frac{1}{2}} = \mathbf{1} \oplus \mathbf{0}$.
	The key observation is that each spontaneous jump operator causes transitions \emph{between} the Dicke subspaces $\mathbf{S}_j$, while the unitary dynamics and cavity decay move the state vector around \emph{within} these subspaces. By organizing $\CH$ into a collection of Dicke subspaces, we can evolve the state vector in only the relevant subspaces, and introduce new subspaces only when we apply a spontaneous jump operator.\par
	The number of subspaces $\mathbf{S}_j$ required doubles for each spontaneously emitted photon, so although we initially avoid using the entire $2^N$-dimensional Hilbert space, the effective Hilbert space still becomes prohibitively large after only a few scattering events. We therefore restrict our simulations to the first five spontaneously emitted photons. In our simulations, trajectories that spontaneously emit all five of these photons do not undergo any further dissipation and are simply evolved unitarily to the end of the simulation. We have chosen parameters such that the number of such trajectories is never more than 10\% of the total, which can introduce an error in $\abs{F}$ of at most 0.1 at late times. There is no such restriction on the number of photons that may be lost by cavity decay.\par
	Our simulations assume that dissipation enters the dynamics only through the $S_x^2$ evolution. An additional source of dissipation is the possibility of photon loss during the controlled phase operation.  The latter effect manifests itself as an overall reduction in the contrast of the interferometric measurement, which can be independently calibrated.  Nevertheless, the interference contrast should be kept of order unity to ensure that the measurement can be performed with good signal-to-noise.  The corresponding requirement on the cooperativity is derived in the following section.\par
	
\section{Conditions on the Cooperativity}
	Here we derive the requirements on the cavity cooperativity $\eta$ that are specified in the main text. Namely, we derive the estimated maximum controlled phase angle $\phi_{\text{max}} \sim \sqrt{\eta / N}$ and the minimum cooperativity $\eta \gtrsim (k \ln N)^2$ required to observe chaotic timescales.\par
\begin{figure}
	\includegraphics[width=.15\textwidth]{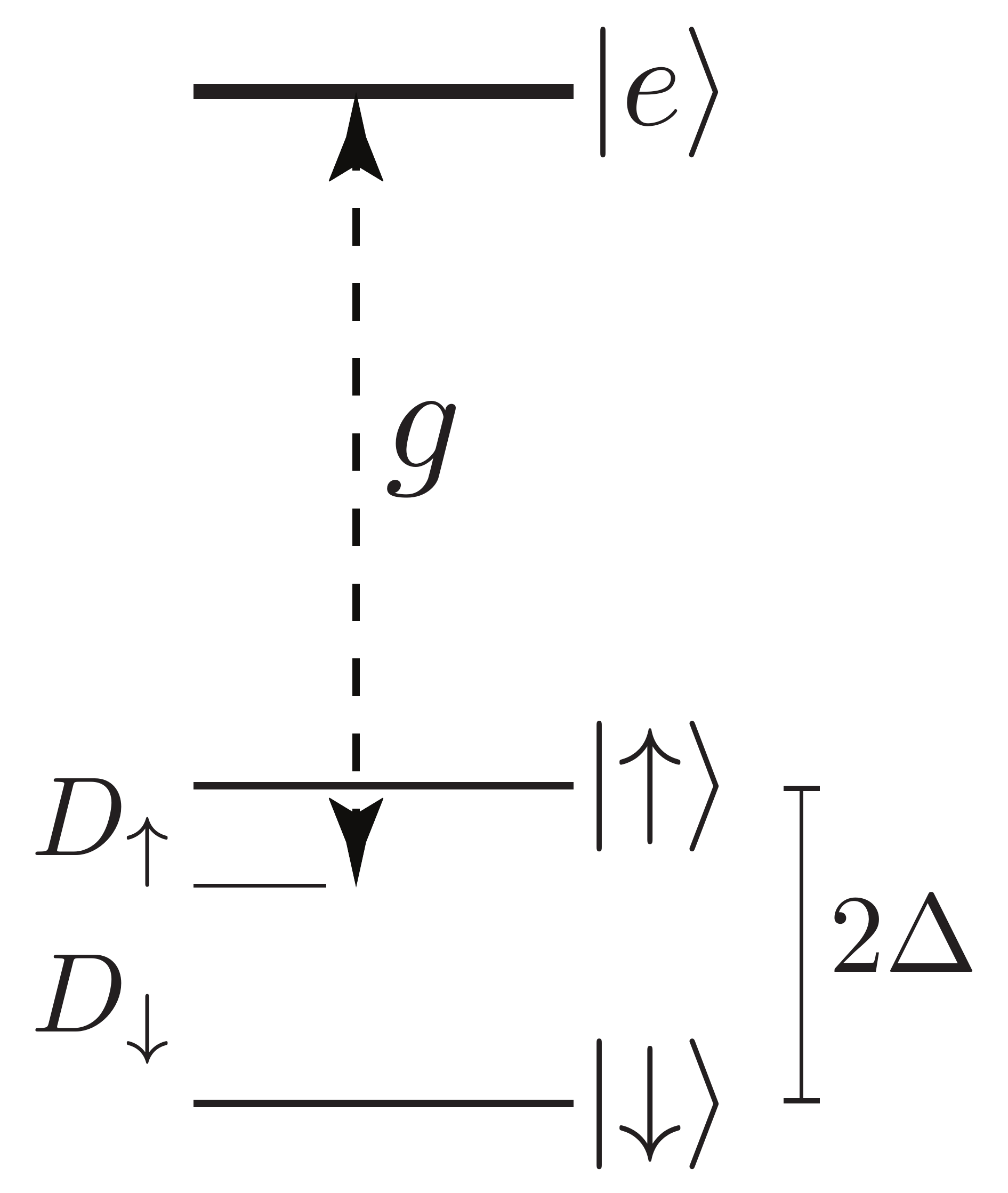}
	\caption{Level scheme for the controlled phase operation assuming uniform cavity coupling $g$. The ground states $\ket{\uparrow}_i, \ket{\downarrow}_i$ could, for instance, be taken to be Zeeman levels from the $F = 1$ and $F = 2$ hyperfine ground state manifolds in $^{87}$Rb.}
	\label{fig:cRotSetup}
	\end{figure}

	The controlled phase operation can be performed by coherently converting the control qubit state into a cavity photon state via stimulated Raman adiabatic passage \cite{Jiang}. If we place the cavity resonance frequency at detunings $D_{\uparrow}, D_{\downarrow}$ from the two ensemble ground states $\ket{\uparrow}_i, \ket{\downarrow}_i$ as shown in Fig. \ref{fig:cRotSetup}, the interactions between the atoms and the cavity result in the effective Hamiltonian \cite{SchleierSmith10a}
	\begin{equation}
		H = \xi \ \adj{a} a S_z,
	\end{equation}
	where $a$ is the cavity mode annihilation operator, $\xi = 2 g^2 / \Delta z$, and we have defined the detuning parameter $z = - D_{\uparrow} D_{\downarrow} / \Delta^2 \leq 1$, where $D_{\downarrow} - D_{\uparrow} = 2 \Delta$. This Hamiltonian applies a rotation of angle $\phi = \xi t$ about the $\hat{z}$-axis if there is a photon in the cavity, and applies the identity if there is not.\par
	The controlled rotation operation fails, however, if we lose the photon due to cavity leakage or spontaneous emission. To ensure a low probability of failure, we therefore require the mean number of photons lost through both processes to be less than 1:
	\begin{equation}
	\label{eq:FailRate}
		N \Gamma_{\text{sc}} t + \kappa t \lesssim 1
	\end{equation}
	where $\Gamma_{\text{sc}} = \Gamma g^2 (2 - z) / \Delta^2 z^2$ is the rate at which photons are lost due to spontaneous emission.\par
	The detuning parameter $z$ determines which dissipation path dominates. Taking $z \rightarrow 0$ implies a small detuning from one of the two ground state transitions, which increases the probability of spontaneous emission. On the other hand, taking $z \rightarrow -\infty$ implies large detuning, for which cavity decay dominates. To find the optimal detuning $z$, we minimize the number of scattered photons assuming fixed absolute rotation angle $\abs{\phi} = \abs{\xi} t$. If we assume the cavity mode lies in between the two ground states $\ket{\uparrow}_i, \ket{\downarrow}_i$ then we have a positive detuning parameter $0 \leq z \leq 1$, and the number of scattered photons is
	\begin{equation}
		N \Gamma_{\text{sc}} t + \kappa t = \left[ \frac{2-z}{z} + \frac{\kappa \Delta^2}{N \Gamma g^2} z \right] \frac{N \Gamma}{2 \Delta} \abs{\phi}.
	\end{equation}
	This is minimized by choosing
	\begin{equation}
		\label{eq:zOpt}
		z_{\text{opt}} = \sqrt{2 N \eta} \left( \frac{\Gamma}{2 \Delta} \right),
	\end{equation}
	where the cavity cooperativity is $\eta = 4 g^2 / \kappa \Gamma$. Substituting Eq. \ref{eq:zOpt} back into Eq. \ref{eq:FailRate}, we see that the achievable controlled rotation angles are limited to:
	\begin{equation}
		\phi \lesssim \phi_{\text{max}} = \sqrt{\frac{\eta}{8 N}}.
	\end{equation}
	Note that the optimum detuning parameter $z_{\text{opt}}$ is accessible only for large ground-state splitting $2 \Delta \geq \sqrt{2 N \eta} \ \Gamma$, which ensures $z_{\text{opt}}\le 1$.  If we choose the ground-state levels to be taken from the $F=1$ and $F=2$ hyperfine manifolds of $^{87}$Rb, the splitting $2\Delta \approx 10^3\Gamma$ is sufficiently large for the above analysis to hold for collective cooperativities as high as $N\eta \sim 10^5$.\par

	The success probability of the controlled phase operation decays with increasing rotation angle $\phi$ as $e^{-\phi/\phi_{\text{max}}}$. The result of imperfect controlled rotation operations is a corresponding reduction in the overall contrast of the $F(t)$ signal. The figures in the main text do not include this loss of contrast, since one can easily compensate for the effect by rescaling such that $F(0) = 1$.\par
	We now turn to the requirement on cooperativity $\eta$ necessary to observe chaotic timescales in $F(t)$. Following the arguments in Ref. \cite{haake}, we require order $\ln N$ kicks in order to observe the onset of classical chaos. During this time, cavity decay and spontaneous emission will act on the ensemble and reduce the fidelity of $F$. The relative strength of these two decay processes is controlled by the cavity detuning parameter $d$ which we should pick so as to minimize the combined effect of both decay channels.\par
	To this end, we first estimate the relative importance of the two decay channels by considering how long it takes each channel to completely decohere an initially pure state. Following arguments in Ref. \cite{Davis15}, smearing generated by the relaxation operator $L_{\kappa}$ causes the spin variance in directions perpendicular to $\hat{x}$ to grow like $\Delta S^2 \sim N^2 \gamma t$. As a result, the cavity will fully dephase the collective spin after a time $\gamma t \sim 1$. Simultaneously, spontaneous emission events produce random spin-flips which will cause complete decoherence of each spin after a time $\mu t \sim 1$. We therefore presume that, given equal rates $\gamma \sim \mu$, the two decay channels are roughly equally destructive. Setting $\gamma = \mu$ requires picking a detuning
	\begin{equation}
		d_{\text{opt}} \approx \sqrt{8 \eta}
	\end{equation}
	for $\eta > 1$.\par
	For the dynamics to remain coherent, we restrict the evolution time so that the Wigner function spreads under cavity decay by less than the width of the initial spin coherent state: $\Delta S^2 \sim N^2 \gamma t \lesssim N$. Note that since $\gamma \sim \mu$, this condition is equivalent to restricting the evolution to the first spontaneously emitted photon. Fixing the minimum number of kicks necessary to observe classical chaos $N \chi t / k \sim \ln N$ and using the detuning $d_{\text{opt}}$, this leads to an estimate for the minimum cooperativity required in order to observe chaos before decoherence destroys the dynamics:
	\begin{equation}
		\eta \gtrsim \left( \frac{k}{2} \ln N \right)^2.
	\end{equation}
	At an experimentally accessible cooperativity $\eta \sim 100$, for $k=3$, this allows for observing the onset of chaos with up to $N \sim 10^3$ atoms.\par

\end{document}